\documentclass[twocolumn,preprintnumbers,prd,superscriptaddress,nofootinbib,floatfix,showpacs,showkeys]{revtex4-1}

\usepackage{amsmath}
\usepackage{amsfonts}
\usepackage{amssymb}
\usepackage{graphicx}
\usepackage{color}
\usepackage{hyperref}
\usepackage{booktabs}
\usepackage{hyperref}
\usepackage{cleveref}
\usepackage{braket}
\usepackage{mathtools}
\usepackage[rightcaption]{sidecap}
\usepackage{subfigure}
\usepackage{soul}
\usepackage{enumitem}

\definecolor{vividviolet}{rgb}{0.62, 0.0, 1.0}
\definecolor{amaranth}{rgb}{0.9, 0.17, 0.31}
\definecolor{palatinateblue}{rgb}{0.15, 0.23, 0.89}
\definecolor{brightpink}{rgb}{1.0, 0.0, 0.5}
\definecolor{cornflowerblue}{rgb}{0.39, 0.58, 0.93}
\definecolor{deepcarminepink}{rgb}{0.94, 0.19, 0.22}
\definecolor{radicalred}{rgb}{1.0, 0.21, 0.37}
\hypersetup{ linktoc=all,
    colorlinks, linkcolor={palatinateblue},
    citecolor={brightpink}, urlcolor={amaranth}
}

\newcommand{\be}{\begin{equation}}
\newcommand{\ee}{\end{equation}}
\newcommand{\bs}{\begin{split}}
\newcommand{\bea}{\begin{eqnarray}}
\newcommand{\eea}{\end{eqnarray}}

\newcommand{\bes}{\begin{subequations}}
\newcommand{\ees}{\end{subequations}}

\begin{document}

\title{Quantum communication through a partially reflecting accelerating mirror}

\author{Michael R. R. Good}
\email{michael.good@nu.edu.kz}
\affiliation{Department of Physics, Nazarbayev University,
Kabanbay Batyr Ave 53, Nur-Sultan, 010000, Kazakhstan.}
\affiliation{Energetic Cosmos Laboratory, Nazarbayev University,
Kabanbay Batyr Ave 53, Nur-Sultan, 010000, Kazakhstan.}

\author{Alessio Lapponi}
\email{alessio.lapponi@studenti.unicam.it}
\affiliation{Divisione di Fisica, Universit\`a di Camerino, Via Madonna delle Carceri, 9, Camerino, 62032, Italy.}

\author{Orlando Luongo}
\email{orlando.luongo@unicam.it}
\affiliation{Divisione di Fisica, Universit\`a di Camerino, Via Madonna delle Carceri, 9, Camerino, 62032, Italy.}
\affiliation{Dipartimento di Matematica, Universit\`a di Pisa, Largo B. Pontecorvo 5, Pisa, 56127, Italy.}
\affiliation{Institute of Experimental and Theoretical Physics, Al-Farabi Kazakh National University, Almaty 050040, Kazakhstan.}

\author{Stefano Mancini}
\email{stefano.mancini@unicam.it}
\affiliation{Divisione di Fisica, Universit\`a di Camerino, Via Madonna delle Carceri, 9, Camerino, 62032, Italy.}
\affiliation{Istituto Nazionale di Fisica Nucleare (INFN), Sezione di Perugia, Perugia, 06123, Italy.}

\begin{abstract}
Motivated by the fact that the null-shell of a collapsing black hole can be described by a perfectly reflecting accelerating mirror, we investigate an extension of this model to mirror semi-transparency and derive a general implicit expression for the corresponding Bogoliubov coefficients. Then, we turn this into an explicit analytical form by focusing on mirrors that are accelerated via an impulsive force. From the so-obtained Bogoliubov coefficients we derive the particle production. Finally, we realize the field coming from left-past spacetime region, passing through the semitransparent moving mirror and ending up to right-future spacetime region as undergoing the action of a Gaussian quantum channel.
We study the transmission and noisy generation properties of this channel, relating them to the Bogoliubov coefficients of the mirror's motion, through which we evaluate capacities in transmitting classical and quantum information.
\end{abstract}



\maketitle
\tableofcontents

\section{Introduction}

The dynamical Casimir effect \cite{moore1970quantum} is the general model encompassing the gravitational analog model of scalar particle creation by a single perfectly reflecting moving mirror \cite{DeWitt:1975ys,Davies:1976hi,Davies:1977yv}. The usual approach to the analog is to assume a prescribed trajectory that fulfills given physical requirements and compute the resulting radiative measures \cite{Good:2016oey,Walker_1982,walker1985particle,Good:2020fjz,Akal:2020twv,Good:2020byh}. A key theoretical success of the dynamical Casimir effect has been the demonstration that accelerated point mirrors disturb the quantum vacuum via a non-zero Bogoliubov transformation and renormalized stress tensor, resulting in principal outputs: particle production, energy flux and entanglement (see e.g. \cite{Bianchi:2014qua, Bianchi:2014vea, Good:2017ddq, Romualdo:2019eur, Good:2019tnf, Cong:2018vqx, Lee:2019adw}). To reconcile the usual divergent stress tensor, point-splitting regularization is used to construct meaningful finite results consistent with particle production (e.g. \cite{Good:2020uff,Stargen:2016euf}). In this prescription, it is found that the particle production and energy flux are a result of the mirror’s acceleration and jerk, respectively \cite{Fabbri,Birrell:1982ix}. The entropy associated with the moving mirror has motivated investigations into thermodynamic puzzles (see e.g. \cite{Davies:1982cn,Helfer:2000fg}) and quantum information issues (see e.g. \cite{Chen:2017lum,Good:2018aer,Giulio}). Efforts are underway to directly\footnote{Superconducting quantum interference device can act as moving mirrors whereas the dynamical Casimir effect can be measured in the case of a Bose-Einstein condensates, see the review \cite{Dodonov:2020eto}.} measure moving mirror radiation \cite{Chen:2020sir,Chen:2015bcg}.

Recently, perfectly reflecting mirror solutions in $(1+1)$-dimensions have been found that demonstrate unexpected resemblances to strong gravitational systems in $(3+1)$-dimensions. Particularly, $(1+1)D$ mirrors could emulate the radiation provided by accelerating boundaries in $(3+1)D$ in terms of particle production and radiated energy. A typical example of the emulated $(3+1)D$ radiation is given by objects undergoing a gravitational collapse into a black hole, leading to Hawking radiation. Currently, analogy between  mirrors and well-known spacetimes, e.g.\ Schwarzschild \cite{Good:2016oey}, Reissner-Nordstr\"om \cite{good2020particle}, Kerr \cite{Good:2020fjz}, and de Sitter/AdS have been found \cite{Good:2020byh}.

Even if the apparent issue related to different dimensions seems to occur, this analogy is predictive and shows the goodness of mirrors in $(1+1)D$ with particular trajectories in describing such physical cases.

Non-thermal or quasi-thermal perfectly reflecting solutions closely characterize other well-known curved spacetime end-states, including extremal black holes (asymptotic uniformly accelerated mirrors \cite{Liberati:2000sq,good2020extreme,Good:2020fjz,Rothman:2000mm,Foo:2020bmv}), black hole remnants (asymptotic constant-velocity mirrors \cite{Good:2016atu,Good:2018ell,Good:2018zmx,Myrzakul:2018bhy,Good:2015nja,Good:2016yht}) and complete black hole evaporation (asymptotic zero-velocity mirrors \cite{Walker_1982, Good:2019tnf,GoodMPLA,Good:2017kjr,B,Good:2017ddq,Good:2018aer}). However, it is worth saying that the reduction from a $(3+1)$-dimensional spacetime to a $(1+1)$-dimensional spacetime does not yield a conformally invariant action. For example, starting from a massless $(3+1)$-dimensional theory in the curved background of spherically symmetric Schwarzschild geometry, transversal (angular) momenta will effectively induce a non-zero mass term in the reduced $(1+1)$-dimensional theory. Hence, the $(1+1)D$ spacetime we are going to deal with does not aim to represent the $(3+1)D$ system\footnote{In the mirror framework, one direct and precise connection that holds in both dimensional contexts, $(3+1)D$ and $(1+1)D$, is the \emph{Lorentz invariant power} as demonstrated in Ref.  \cite{Zhakenuly:2021pfm}}.

Remarkably, generalizing Bogolyubov coefficients in $(3+1)$-dimensional theory has so far been intractable. Consequently, attempts toward particle production analysis in $(3+1)D$ spacetime is beyond the scope of this paper.

Despite impressive progress over the last half-century \cite{Dodonov:2020eto}, the moving mirror model is still evolving. The extension to realistic conditions for partially transmitting mirrors has had success in generalizing the specialized case of perfectly reflecting mirrors which often posses infrared divergences \cite{Frolov:1999bi,Nicolaevici:1999ga,Nicolaevici_2001,Nicolaevici:2011zza,Fosco:2017jjf,Nicolaevici:2010zz}. Semi-transparent mirrors can also be used to simulate a null-shell collapse to form a black hole and provide new insights in determining the physics of particle production \cite{Calogeracos,Nicolaevici:2009zz}.

Considering semitransparent mirrors, we provide a more realistic case for the dynamical Casimir effect since, in real mirrors, perfect reflection is only an approximation valid for a small range of frequencies. For the black hole-mirror analogy, perfect reflection models the regularity condition at the center of the collapsing ball $r=0$, where $r$ is the radial coordinate in $(3+1)D$ spacetime\footnote{{We remind that the analogy occurs between $(1+1)D$ mirrors and $(3+1)D$ collapsing balls or shells}}. This condition says that the field vanishes at $r=0$ because no field can exist behind $r<0$, as the coordinate itself is defined only for $r\geq 0$. The semitransparency of the mirror stresses out this condition: this may seem unphysical. However, in important and interesting contexts where it becomes impossible to impose regularity, say e.g. 4-dimensional Schwarzschild spacetime with Eddington-Finkelstein coordinates such that $r=0$ is a spacelike curvature singularity (see e.g. Eq. (5.137) of Ref. \cite{Fabbri}), perfect reflection might indeed need to be relaxed.  For this reason, there is good physical motivation to study semi-transparent moving mirrors with respect to black hole radiation.

Another intriguing issue is the interplay between mirrors and quantum information theory. A mirror can be seen as a fundamental tool to model a quantum communication channel.
Since the relation between input and output modes through a mirror is linear, it actually realizes a bosonic Gaussian channel. In this perspective, the mirror is however always considered at rest. Only recently quantum channels arising from the reflection of a one-mode bosonic input upon a perfectly reflecting moving mirror have been characterized \cite{Giulio}.
It seems then quite natural to investigate the information transmission capabilities of quantum channels arising in the broader context of semitransparent moving mirrors. This would allow us to also explore the information capabilities across the null shell of a collapsing black hole.

In this paper we investigate the particle production from a semitransparent moving mirror,  obtaining analytical expressions of Bogoliubov coefficients. To do this, we consider a very short acceleration period compared with the wavelength of the produced particles, namely we focus on impulsive accelerated semitransparent mirrors. Consequently, we obtain a finite spectrum of the radiated particles. Then, we investigate the transmission of a signal, carried by the field, through a semitransparent moving mirror.
The above mentioned spectrum permits to understand if the mirror motion can improve the quality of the signal transmission, or if it only creates additional noise, compared with the static case.
Actually, we shall realize the field coming from left-past spacetime region, passing through the semitransparent moving mirror and ending up to right-future spacetime region as undergoing the action of a Gaussian quantum channel,
obtaining an average transmission coefficient, $\tau$, and an average number of noisy particles created,  $\overline{n}$. For a mirror with a short acceleration period we find $\tau<1$ and $\overline{n}=0$, yielding a beam splitter bosonic channel. Therefore, an exact expression of the classical and quantum capacity is provided. The most interesting property arising from this line of investigation is that, for each frequency of the input signal, $\tau$ is maximized when the final speed of the mirror is equal to a critical speed, which is different from the speed of light.

The paper is organized in the following: in Sec.~\ref{Sec2} we provide the general expressions for Bogoliubov coefficients for semitransparent moving mirrors, assuming the trajectory of the mirror starting from time-like past and ending at time-like future in proper null coordinates. in Sec.~\ref{Sec3} we focus on trajectories which have a finite acceleration period leading to analytic expressions when this period is very small. In Sec.~\ref{Sec4} we show that the transmission of a signal through a semitransparent moving mirror corresponds to the transmission through a bosonic Gaussian quantum channel, following the same procedure used in Ref.  \cite{Giulio}.  Finally, in Sec. \ref{Sec5} we provide an exact expression for the classical and quantum capacity of the quantum channel created by an impulsive accelerated mirror. Throughout we use natural units, namely $\hbar=c=1$.

\color{black}
\section{Bogoliubov coefficients for moving mirrors with proper null coordinates}\label{Sec2}

In this section we propose a general, thought implicit, expression for the Bogoliubov coefficients relating input (in) and output (out) modes in the presence of a semitransparent moving mirror. These coefficients give information about the spectrum of particles produced by the mirror and
will be at the heart of the communication properties of the mirror.

\begin{figure}
    \centering
    \includegraphics[scale=0.5]{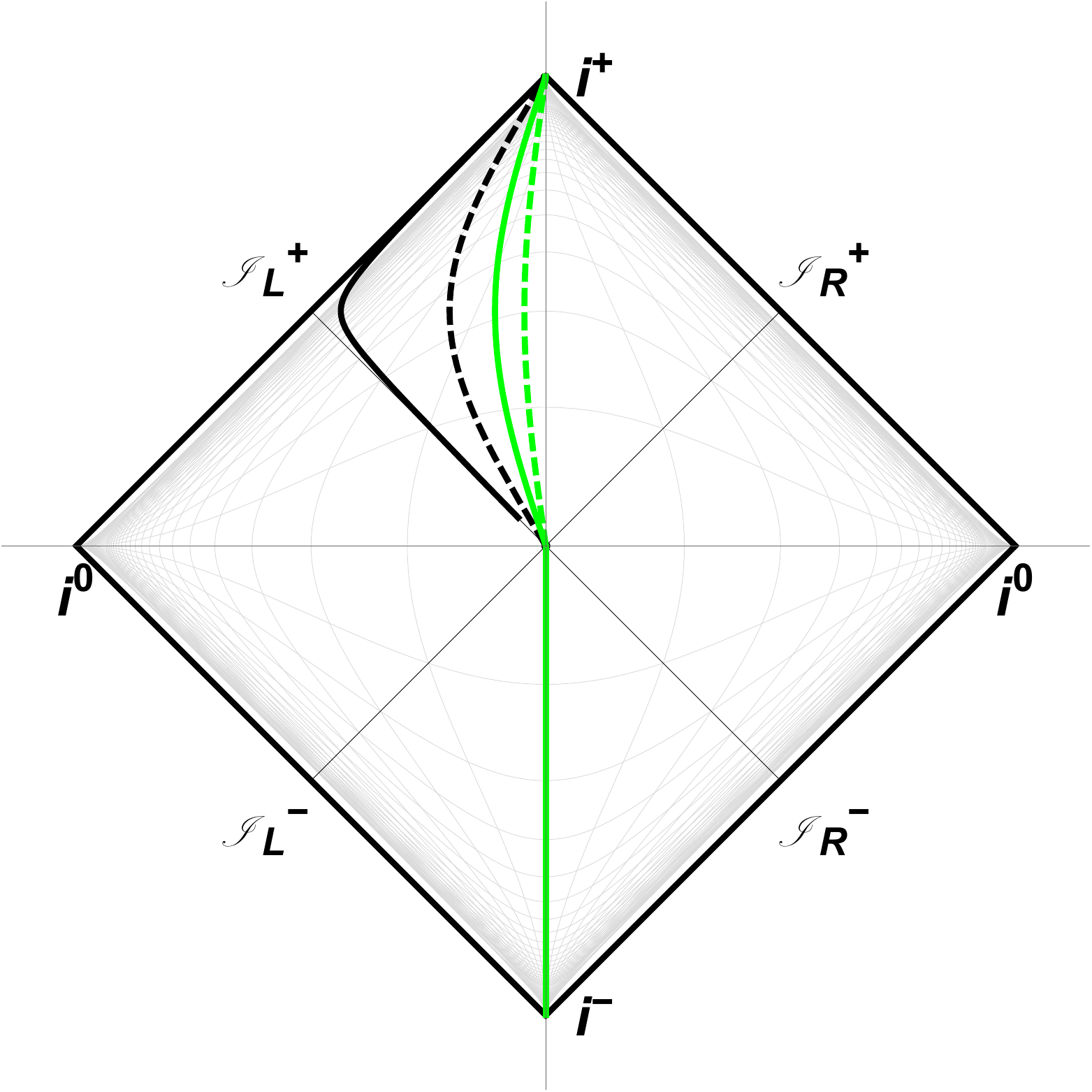}
    \caption{Penrose diagram showing the trajectories of high acceleration mirrors for different values of the parameter $\nu$, defined in Sec.~\ref{Sec3}. The trajectory is like the one described by Eq.~\eqref{pacc} with $u_0$ very small. In particular, $\nu=1.5$ for the green dashed line, $\nu=2$ for the green line, $\nu=4$ for the black dashed line and $\nu=100$ for the black line. We can imagine the infinitesimal acceleration period to be in a neighborhood of $t=0$.}
    \label{penrose}
\end{figure}

We work on a $(1+1)D$ spacetime, which can be compactly portrayed through Penrose diagrams, as in Fig.~\ref{penrose}.

As stressed in the introduction, the $(3+1)$-dimensional case is more realistic, even if very harsh to study analytically. Nevertheless, the $(1+1)D$ results provide a suitable matching with theoretical expectations in $(3+1)D$ spacetimes, e.g. recovering the Hawking radiation and/or obtaining the dynamical Casimir effect when the mirror is very large. Consequently, the next results are thought to hold a relevant guidance for $(3+1)D$ spacetimes.

There, $i^-$ and $i^+$ represent time-like past and future infinities, respectively. The null surfaces $\mathcal{J}_{R/L}^\pm$ are instead the boundaries of the Penrose diagram. Since only massless scalar particles will be considered as input and output, the input mode should necessarily come from a past null-like surface $\mathcal{J}_{R/L}^-$, whereas the output mode should end up at a future null-like surface $\mathcal{J}_{R/L}^{\pm}$. To this end, we simply introduce the null coordinates $u=t-x$ and $v=t+x$. The trajectory of a mirror is usually expressed via null coordinates through the function $p(u)\coloneqq v_{mirror}$ and its inverse $f(v)\coloneqq u_{mirror}$, e.g. \cite{Davies:1976hi}. To guarantee the mirror does not evolve faster than light both $p(u)$ and $f(v)$ are increasing monotonic functions.

As anticipated, we only consider a massless scalar field $\Phi$ (since it describes the vast majority of radiation fields) without self-interaction. The Lagrangian density for this field interacting with a static mirror at the position $x_m$ is described by \cite{Calogeracos}:
\begin{equation}
    \mathcal{L}=\frac{1}{2}\partial_\mu\Phi\partial^\mu\Phi+\eta\Phi\delta(x-x_m).
\end{equation}
From this, one can obtain the following reflection and transmission amplitudes:
\begin{equation}
r(\omega)=-\frac{i\eta}{\omega+i\eta}\,,\hspace{1 cm} s(\omega)=\frac{\omega}{\omega+i\eta},
\end{equation}
where $\omega$ is the frequency of the reflected/transmitted mode.

The field $\Phi$ can be expanded as:
\begin{equation}
 \Phi=\sum_{J=L,R}\int_0^\infty\left(\phi_\omega^Ja_\omega^J+\phi_\omega^{J*}a_\omega^{J\dagger}\right)d\omega,
\end{equation}
where $\phi^R_\omega$ (resp. $\phi^L_\omega$) is the input mode with frequency $\omega$ incoming from the right, i.e. $\mathcal{J}_R^-$ (resp. left, i.e. $\mathcal{J}_L^-$) and $a_\omega^R$ (resp. $a_\omega^L$) is the corresponding annihilation operator. Considering the boundary condition given by the mirror at the position $x=x_m$, the modes $\phi_\omega^R$ and $\phi_\omega^L$ can be written, respectively, as:
\begin{align}\label{semitstaticR}
\phi_{\omega}^{R}(u,v)&=\frac{1}{\sqrt{4\pi|\omega|}}\left(s(\omega)e^{-i\omega v}\theta(u-v+2x_m)\right.\notag\\
&\left.+\left(e^{-i\omega v}+r(\omega)e^{-i\omega u}\right)\theta(v-u-2x_m)\right),\\
\phi_{\omega}^L(u,v)&=\frac{1}{\sqrt{4\pi|\omega|}}\left(s(\omega)e^{-i\omega u}\theta(v-u-2x_m)\right.
\notag\\&\left.+\left(e^{-i\omega u}+r(\omega)e^{-i\omega v}\right)\theta(u-v+2x_m)\right).
\label{semitstaticL}
\end{align}

The expressions for the input modes \eqref{semitstaticR} and \eqref{semitstaticL} are valid when the mirror is static. When the mirror moves along a trajectory $x_m(t)$ also the amplitudes $r(\omega)$ and $s(\omega)$ change in time (see Appendix A of \cite{Nicolaevici:2009zz}). Hence, there is a great mathematical complication, since the boundary condition between the left side and right side of the mirror becomes time-dependent. To overcome this problem, we resort to the strategy used in Ref.~\cite{C}. Namely, we put ourselves in the mirror frame, using proper coordinates. The proper distance from the mirror is indicated by $\rho$ and the proper time by $\tau$. From them, the proper null coordinates are defined as $\overline{u}\coloneqq\tau-\rho$ and $\overline{v}=\tau+\rho$. The proper null coordinates $\overline{u}$ and $\overline{v}$ could also be written in terms of the external null coordinates $u=t-x$ and $v=t+x$ through $\overline{u}(u)$ and $\overline{v}(v)$ which depend on the trajectory of the mirror (specified by $p(u)$ and $f(v)$). This dependence comes from the metric conservation (see Eq.~(20) of \cite{Obadia:2001hj}):
\begin{equation}
   ds^2=d\overline{u}^2=d\overline{v}^2=\partial_up(u)du^2=\partial_vf(v)dv^2,
\end{equation}
from which one obtains:
\begin{equation}
	\frac{d\overline{u}(u)}{du}=\sqrt{\partial_up(u)},\hspace{1 cm}\frac{d\overline{v}(v)}{dv}=\sqrt{\partial_vf(v)}.
\end{equation}
Finally, we define the input modes in proper coordinates as $g_\omega^R$ and $g_\omega^L$.

In the mirror frame, the mirror is obviously static. Hence, the modes $g_\omega^R$ and $g_\omega^L$ could be written analogously to the modes $\phi_\omega^R$ and $\phi_\omega^L$ in the static case, i.e. like Eqs.~\eqref{semitstaticR} and \eqref{semitstaticL}:
\begin{align}\label{propermodeR}
&g_{\omega}^{R}(\overline{u},\overline{v})=\frac{1}{\sqrt{4\pi|\omega|}}\notag\\
&\times\left(s(\omega)e^{-i\omega\overline{v}}\theta(\overline{u}-\overline{v})+\left(e^{-i\omega\overline{v}}+r(\omega)e^{-i\omega\overline{u}}\right)\theta(\overline{v}-\overline{u})\right),\\
&g_{\omega}^{L}(\overline{u},\overline{v})=\frac{1}{\sqrt{4\pi|\omega|}}\notag\\
&\times\left(s(\omega)e^{-i\omega\overline{u}}\theta(\overline{v}-\overline{u})+\left(e^{-i\omega\overline{u}}+r(\omega)e^{-i\omega\overline{v}}\right)\theta(\overline{u}-\overline{v})\right).
\label{propermodeL}
\end{align}

From now on, we consider only time-like trajectories for the mirror, i.e. we consider trajectories starting at $i^-$ and ending up at $i^+$, referring to the Penrose diagram in Fig.~\ref{penrose}. In that case, both the modes in external coordinates $\{\phi_\omega^R\}_v$ and the ones in proper coordinates $\{g_\omega^R\}_{\overline{v}}$ form a complete set of input modes incoming from $\mathcal{J}_{R}^{-}$. Analogously, both the sets $\{\phi_\omega^L\}_u$ and $\{g_\omega^L\}_{\overline{u}}$ are complete sets of modes incoming from $\mathcal{J}_L^{-}$. As a consequence, the modes in external coordinates $\phi_\omega^J$ are related to the ones in proper coordinates through the following Bogoliubov transformation \cite{C}:
\begin{equation}\label{inmodes}
\phi^{J}_{\omega}=\int_{-\infty}^{+\infty}\chi(\omega')\left(\phi_{\omega}^{J},g_{\omega'}^{J}\right)g_{\omega'}^{J}d\omega',
\end{equation}
where $\chi$ is the sign function.

Using Eq.~\eqref{inmodes} one may obtain a general expression for the input modes in external coordinates $\phi_\omega^J$. Let us turn our attention to the scalar product $\left(\phi_{\omega}^{R},g_{\omega'}^{R}\right)$. To single out a convenient integration surface for the integration, we select $\mathcal{J}_R^-$, since here $\phi_{\omega}^{R}=\frac{1}{\sqrt{4\pi|\omega|}}e^{-i\omega v}$ and $g_\omega^R=\frac{1}{\sqrt{4\pi|\omega|}}e^{-i\omega \overline{v}(v)}$. So, the scalar product becomes
\begin{equation}\label{scprR}
\left(\phi_\omega^R,g_{\omega'}^R\right)=-i\int_{-\infty}^{+\infty}\left((\partial_vg_{\omega'}^{R*})\phi_{\omega}^{R}-g_{\omega'}^{R*}\partial_v\phi_{\omega}^R\right)dv.
\end{equation}
Using the fact that these modes vanish for $v\rightarrow\pm\infty$ we can integrate Eq.~\eqref{scprR} by parts, simplifying the scalar product to
\begin{align}
\left(\phi_{\omega}^{R},g_{\omega'}^{R}\right)&=2i\int_{-\infty}^{+\infty}g_{\omega'}^{R}\partial_v\phi_{\omega}^Rdv\notag\\
&=\frac{\omega}{2\pi\sqrt{|\omega||\omega'|}}\int_{-\infty}^{+\infty}e^{-i\left(\omega v-\omega'\overline{v}(v)\right)}dv.
\end{align}
The same thing is done for the scalar product $\left(g_{\omega'}^{L},\phi_{\omega}^{L}\right)$ integrating on the surface $\mathcal{J}_L^-$ and obtaining
\begin{align}
\left(\phi_{\omega}^{L},g_{\omega'}^{L}\right)&=2i\int_{-\infty}^{+\infty}g_{\omega'}^{L*}\partial_v\phi_{\omega}^Ldv\notag\\
&=\frac{\omega}{2\pi\sqrt{|\omega||\omega'|}}\int_{-\infty}^{+\infty}e^{-i\left(\omega u-\omega'\overline{u}(u)\right)}du.
\end{align}
From now on, we also consider a detector positioned on the right of the mirror. Hence, the terms of $g_\omega^J$ (from Eqs.~\eqref{propermodeR} and \eqref{propermodeL}) proportional to $\theta(\overline{u}-\overline{v})$ are neglected. For the mode coming from the right of the mirror $\phi_\omega^R$, applying some contour integration in the variable $\omega'$ and using the fact that $\overline{u}(u)$ and $\overline{v}(v)$ are increasing monotonic functions (to guarantee that the mirror speed is not faster than the speed of light), we obtain
\begin{equation}
\phi^{R}_{\omega}=\frac{1}{\sqrt{4\pi\omega}}e^{-i\omega v}+\phi^{refl}_{\omega,R}(u),
\end{equation}
where $\phi^{refl}_{\omega}(u)$ is the reflected wave of the input mode:
\begin{align}
\phi^{refl}_{\omega}(u)&=\sqrt{\frac{\omega}{4\pi}}i\int_{-\infty}^{+\infty}\left(\frac{1}{2}\chi(\overline{u}(u)-\overline{v}(v'))\right.\notag\\
&\left.-\theta(\overline{u}(u)-\overline{v}(v'))e^{-\eta(\overline{u}(u)-\overline{v}(v')}\right)e^{-i\omega v'}dv'.
\end{align}
With a similar calculation, we can obtain $\phi_{\omega}^{L}$  as
\begin{align}
\phi_{\omega}^{L}&=\sqrt{\frac{\omega}{4\pi}}(-i)\int_{-\infty}^{+\infty}\theta(u-u')e^{-\eta\left(\overline{u}(u)-\overline{u}(u')\right)}e^{-i\omega u'}du'\nonumber\\
&=\sqrt{\frac{\omega}{4\pi}}(-i)\int_{-\infty}^{u}e^{-\eta\left(\overline{u}(u)-\overline{u}(u')\right)}e^{-i\omega u'}du'\,.
\end{align}

In the presence of an accelerating boundary, we expect a difference between the input and the output spacetime structure \cite{Hawking:1974sw, Birrell:1982ix,CarlitzWilley}. The output mode (outgoing to the right, i.e. $\mathcal{J}_R^+$) can be written in terms of the input ones through the following Bogoliubov transformation:
\begin{equation}
    \phi_\omega^{out}=\sum_{J=R,L}\int_0^\infty\left(\alpha_{\omega\omega'}^{RJ}\phi_{\omega'}^J+\beta_{\omega\omega'}^{RJ}\phi_{\omega'}^{J*}\right)d\omega',
\end{equation}
where $\alpha_{\omega\omega'}^{RR}=\left(\phi_{\omega}^{out},\phi_{\omega'}^{R}\right)$, $\alpha_{\omega\omega'}^{RL}=\left(\phi_{\omega}^{out},\phi_{\omega'}^{L}\right)$, $\beta_{\omega\omega'}^{RR}=\left(\phi_{\omega}^{out*},\phi_{\omega'}^{R}\right)^*$ and $\beta_{\omega\omega'}^{RL}=\left(\phi_{\omega}^{out*},\phi_{\omega'}^{L}\right)^*$ are the Bogoliubov coefficients we want to compute. We are particularly interested in the $\beta$ Bogoliubov coefficients, since they give us information about the production of the particles by the mirror. Since we consider mirrors with a trajectory ending at $i^+$, $\mathcal{J}^+_R$ is a surface we can use as an integration surface for the scalar product of Bogoliubov coefficients. In fact on this surface we have

\begin{subequations}
\begin{align}
\phi_{\omega}^{out*}&\longrightarrow\frac{1}{\sqrt{4\pi\omega}}e^{i\omega u},\\
\phi_{\omega'}^{R}&\longrightarrow\phi^{refl}_{\omega',R},
\end{align}
\end{subequations}
Applying the derivative of the scalar product to the input modes $\phi_\omega^{J}$ and changing the variable from $u$ to $\overline{u}$ (we define $u(\overline{u})$ as the inverse of $\overline{u}(u)$) we obtain the following general expression for the Bogoliubov coefficients:
\pagebreak
\begin{widetext}
\begin{align}\label{betacoeffR}
&\beta_{\omega\omega'}^{RR}=-\frac{\eta}{2\pi}\sqrt{\frac{\omega'}{\omega}}\int_{-\infty}^{+\infty}\int_{-\infty}^{+\infty}\theta(\overline{u}-\overline{v}(v'))e^{\eta\overline{v}(v')-i\omega'v'}e^{-\eta\overline{u}-i\omega u(\overline{u})}dv'd\overline{u},\\
&\beta_{\omega\omega'}^{RL}=-\frac{\eta}{2\pi}\sqrt{\frac{\omega'}{\omega}}\int_{-\infty}^{+\infty}\int_{-\infty}^{+\infty}\theta(\overline{u}-\overline{u}(u'))e^{\eta\overline{u}(u')-i\omega'u'}e^{-\eta\overline{u}-i\omega u(\overline{u})}du'd\overline{u},\label{bogcoeffL}
\end{align}
\end{widetext}
and
\begin{widetext}
\begin{align}\label{alphaR}
&\alpha_{\omega\omega'}^{RR}=-\frac{\eta}{2\pi}\sqrt{\frac{\omega'}{\omega}}\int_{-\infty}^{+\infty}\int_{-\infty}^{+\infty}\theta(\overline{u}-\overline{v}(v'))e^{\eta\overline{v}(v')+i\omega'v'}e^{-\eta\overline{u}-i\omega u(\overline{u})}dv'd\overline{u},\\
&\label{alphaL}\alpha_{\omega\omega'}^{RL}=\sqrt{\frac{\omega'}{\omega}}\delta(\omega-\omega')-\frac{\eta}{2\pi}\sqrt{\frac{\omega'}{\omega}}\int_{-\infty}^{+\infty}\int_{-\infty}^{+\infty}\theta(\overline{u}-\overline{u}(u'))e^{\eta\overline{u}(u')+i\omega'u'}e^{-\eta\overline{u}-i\omega u(\overline{u})}du'd\overline{u}.
\end{align}
\end{widetext}
Considering Eqs. \eqref{betacoeffR} and \eqref{bogcoeffL}, we have that $N_\omega^R=\int_{0}^{+\infty}\left|\beta_{\omega\omega'}^{RR}\right|^2d\omega'$ is the spectrum of the particles produced due the reflection of the modes at the right of the mirror and $N_\omega^L=\int_{0}^{+\infty}\left|\beta_{\omega\omega'}^{RL}\right|^2d\omega'$ is the spectrum of particles produced due the transmission of the modes at the left of the mirror. The total spectrum of produced particles is therefore given by
\begin{equation}
N_\omega=N_\omega^R+N_\omega^L.
\end{equation}


\section{Approximated trajectories with impulsive acceleration} \label{Sec3}

Once we have found a general expression for the Bogoliubov coefficients, as reported in Eqs. \eqref{betacoeffR}, \eqref{bogcoeffL}, \eqref{alphaR} and \eqref{alphaL}, we aim at finding trajectories for which the Bogoliubov coefficients can be explicitly and analytically computed.
In this Section we show that this task can be accomplished for trajectories corresponding to impulsive acceleration.

First, looking at Eqs.  \eqref{betacoeffR} and \eqref{bogcoeffL}, we notice that finding explicit Bogoliubov coefficients turns out to be non-analytical and quite hard for some physical aspects that we summarize below. \begin{itemize}[noitemsep]
	\item[{\bf 1)}] The trajectories in null comoving coordinates are arguments of exponentials, and the only functions which can be easily analytically computed in an exponential are the linear ones, which corresponds to a non-accelerating mirror, giving trivially the Bogoliubov coefficients $\beta=0$.
	\item[{\bf 2)}] In order to find a non-trivial trajectory we have to use integral functions (such as the Euler gamma), but even if we solve one integral with it, we also need to solve the other. This will be an integral of an exponential multiplied by an integral function, which is almost always not computable analytically.
	\item[{\bf 3)}] In particular for Eq.~\eqref{bogcoeffL} we need to find a function $\overline{u}(u)$ such that the exponential of it and the exponential of its inverse $u(\overline{u})$ is integrable analytically. Even if such a function exists, it is needed also that Eq.~\eqref{betacoeffR} is analytically computable.
	\item[{\bf 4)}] For some trajectories (such as the Carlitz-Willey's and the uniform acceleration trajectory \cite{CarlitzWilley}) we can pass easily from the expressions of the trajectories in null coordinates to $\overline{u}(u)$ and $\overline{v}(v)$. However, for the other non-trivial trajectories which provides an exact solution for the Bogoliubov coefficients in the perfectly reflecting case (Walker-Davies \cite{Walker_1982}, Arctx \cite{B}, Dlogex \cite{Good_2015BirthCry}, Proex \cite{Good:2016yht}, etc.) it is impossible even to find an analytic expression for the trajectories in null coordinates $\overline{u}(u)$ and $\overline{v}(v)$. Reversely, if we succeed finding a function $\overline{u}(u)$, such that we can find an analytic function for the Bogoliubov coefficients, we need to have the respective functions $p(u)$ and $f(v)$ describing the trajectory of a mirror. To this aim, the quantities $p(u)$ and $f(v)$ should be real, without asymptotes, without singularities in their domain, and monotonic in their domain.
\end{itemize}
We thus consider a class of trajectories that will allow us to simplify the treatment.
It consists of trajectories fulfilling the conditions below:
\begin{itemize}[noitemsep]
	\item the mirror is static at $x=0$ for $t<0$,
	\item it begins to accelerate toward its left along a certain trajectory $z(t)$ at $t=0$ until it reaches the time $t_0$, arriving to the point $x_0$,
	\item after a time interval $t_0$, the mirror continues travelling with the same velocity $V$ reached after the acceleration period. Such $V$ must be smaller than $1$, otherwise the proper acceleration of the mirror would become infinite.
\end{itemize}
Such a trajectory can be described though the simplest choice, represented by the following function
\begin{equation}\label{pacc}
p_{tot}(u)=\begin{cases}
u, & \text{if}\;u\le 0,\\
p(u), & \text{if}\;0<u\le u_0,\\
p(u_0)+\nu^{-1}(u-u_0), & \text{if}\;u>u_0.
\end{cases}
\end{equation}
For simplicity, hereafter we refer to the complete trajectories with the subscript ``tot" and with the usual $p(u)$, $f(v)$, $\overline{u}(u)$ and $\overline{v}(v)$ we refer to the trajectory of the mirror only on its period of acceleration. In Eq.~\eqref{pacc} $u_0=t_0-x_0$ quantifies the width of the acceleration period, alongside with $\nu^{-1}:=\partial_u p(u)|_{u_0}$, which ensures the continuity of the derivative on $u_0$\footnote{The continuity of $p_{tot}(u)$ and of its first derivative should be imposed at $u=0$ as well. Hence, the trajectory $p(u)$ should be chosen to satisfy this condition.}. The quantity $\nu$ is related to the final speed of the mirror $V$ through $\nu=\frac{1+V}{1-V}$ and any deviations from the linearity in terms of $\sim u$ would imply that, at asymptotic regimes, the velocity is no longer a constant.

Analogously, we can write the inverse of $p_{tot}(u)$, i.e., $f_{tot}(v)$ by
\begin{equation}\label{facc}
f_{tot}(v)=\begin{cases}
v, & \textrm{if}\;v\le 0,\\
f(v), & \textrm{if}\;0<v\le v_0,\\
f(v_0)+\nu(v-v_0), & \textrm{if}\;v>v_0,
\end{cases}
\end{equation}
where $v_0=p(u_0)$. It is worth noticing that all the quantities derived from $u_0$ ,i.e., $v_0$, $\overline{u}_0=\overline{u}(u_0)$ and $\overline{v}_0=\overline{v}(v_0)$ satisfy the equations describing the trajectories of the mirror ,i.e., $v=p(u)$, $u=f(v)$, $\overline{u}=\overline{v}(v)$ and $\overline{v}=\overline{u}(u)$. As consequence we have: $v_0=p(u_0)$, $u_0=f(v_0)$ and $\overline{u}(u_0)=\overline{v}(v_0)=\overline{u}_0=\overline{v}_0$.\\
\indent In proper coordinates Eq.~\eqref{pacc} and Eq.~\eqref{facc} become:
\begin{equation}\label{trajgenu}
\overline{u}_{tot}(u)=\begin{cases}
u, & \text{if}\;u\le 0,\\
\overline{u}(u), & \text{if}\;0<u\le u_0,\\
\overline{u}(u_0)+\nu^{-1/2}(u-u_0), & \text{if}\;u>u_0,
\end{cases}
\end{equation}
\begin{equation}\label{trajgenv}
\overline{v}_{tot}(v)=\begin{cases}
v, & \text{if}\;v\le 0,\\
\overline{v}(v), & \text{if}\;0<v\le v_0,\\
\overline{v}(v_0)+\sqrt{\nu}(v-v_0), & \text{if}\;v>v_0.
\end{cases}
\end{equation}
We now focus on the $\beta$ Bogoliubov coefficients (in order to obtain the $\alpha$ it is sufficient to switch $\omega'\rightarrow-\omega'$ everywhere except in the external factor $-\frac{\eta}{2\pi}\sqrt{\frac{\omega'}{\omega}}$, and add the term $\sqrt{\frac{\omega'}{\omega}}\delta(\omega-\omega')$ in $\alpha_{\omega\omega'}^{RL}$). For this class of trajectories we obtain, separating the integrals, through the Eq.~\eqref{betacoeffR} and Eq.~\eqref{bogcoeffL} the following

\begin{widetext}
\begin{equation}\label{accrbc}
\begin{split}
\beta_{\omega\omega'}^{RR}=-\frac{\eta}{2\pi}\sqrt{\frac{\omega'}{\omega}}\left[\frac{e^{-i\omega u_0-\eta\overline{u}_0}}{\left(\omega\sqrt{\nu}-i\eta\right)(\omega'+i\eta)}-\frac{1}{(\omega'+i\eta)(\omega+\omega'+i\epsilon)}-\frac{e^{-i\omega u_0-i\omega'v_0}}{\left(\omega\nu+\omega'-i\nu\epsilon\right)\left(\omega\sqrt{\nu}-i\eta\right)}\right.\\\left.+\frac{1}{\eta-i\omega'}\int_{0}^{\overline{u}_0}e^{-\eta\overline{u}-i\omega u(\overline{u})}d\overline{u}+\frac{e^{-i\omega u_0-\eta\overline{u}_0}}{\eta+i\omega\sqrt{\nu}}\int_{0}^{v_0}e^{\eta\overline{v}(v')-i\omega'v'}dv'\right.\\\left.+\int_{0}^{\overline{u}_0}e^{-\eta\overline{u}-i\omega u(\overline{u})}\left(\int_{0}^{v(\overline{u})}e^{\eta\overline{v}(v')-i\omega'v'}dv'\right)d\overline{u}\right];
\end{split}
\end{equation}
\end{widetext}

\begin{widetext}
\begin{equation}\label{acclbg}
\begin{split}
\beta_{\omega\omega'}^{RL}=-\frac{\eta}{2\pi}\sqrt{\frac{\omega'}{\omega}}\left[\frac{e^{-i\omega u_0-\eta\overline{u}_0}}{\left(\omega\sqrt{\nu}-i\eta\right)(\omega'+i\eta)}-\frac{1}{(\omega'+i\eta)(\omega+\omega'+i\epsilon)}-\frac{e^{-i(\omega+\omega')u_0}}{\left(\omega+\omega'-i\epsilon\right)\left(\omega\sqrt{\nu}-i\eta\right)}\right.\\
\left.+\frac{1}{\eta-i\omega'}\int_{0}^{\overline{u}_0}e^{-\eta\overline{u}-i\omega u(\overline{u})}d\overline{u}+\frac{e^{-i\omega u_0-\eta\overline{u}_0}}{\eta+i\sqrt{\nu}\omega}\int_{0}^{u_0}e^{\eta\overline{u}(u')-i\omega'u'}du'\right.\\
\left.+\int_{0}^{\overline{u}_0}e^{-\eta\overline{u}-i\omega u(\overline{u})}\left(\int_{0}^{u(\overline{u})}e^{\eta\overline{u}(u')-i\omega'u'}du'\right)d\overline{u}\right],
\end{split}
\end{equation}
\end{widetext}
where $\epsilon$ is an exponential cutoff in $u$ and $v$, needed in order to make some integrals convergent. For the $\beta$ Bogoliubov coefficients, we can set $\epsilon=0$ without problems. However, for the $\alpha$ Bogoliubov coefficients we cannot neglect $\epsilon$, otherwise a divergence for them occurs for $\omega=\omega'$\footnote{For Eq.~\eqref{alphaL}, considering $\epsilon\ne0$, the Dirac delta can be seen as $\frac{1}{\pi}\frac{\epsilon}{\epsilon^2+(\omega-\omega')^2}$.}.

For computing particle production, we only need the $\beta$ Bogoliubov coefficients. So, limiting to $\epsilon=0$ and looking at Eqs. \eqref{accrbc} and \eqref{acclbg}, we soon notice that we need a strategy to neglect those integrals that are non-analytical.

Since the arguments of all the integrals have no singularities in their integration range, and since for $u_0\rightarrow0$ we have also $v_0\rightarrow0$ and $\overline{u}(u)\rightarrow0$, the first option is to consider $u_0\rightarrow0$, taking the acceleration period so much short to  neglect its contribution. Further, with the recipe $u_0\sim0$, we stress  that, before this period the mirror was at rest, whereas after it the mirror shows a finite velocity.

Since we are minimizing the period in which the mirror accelerates, we could  maximize the acceleration. In this respect, we refer to these mirrors as ``impulsive accelerated mirrors", in which particular physical consequences are expected. In particular, to clarify why we need to maximize the acceleration, let us first consider the well-consolidate  Carlitz-Willey trajectory \cite{CarlitzWilley}. Here, we have $\overline{u}(u)=\frac{2}{k}\left(1-e^{-ku/2}\right)$ and $\overline{v}(v)=\frac{2}{k}\left(1-\sqrt{1-kv}\right)$, where $k$ is intimately related to the mirror acceleration.

Even though we cannot fix the acceleration, since it is not constant for the Carlitz-Willey trajectory, we can fix the parameter $k$ to be arbitrarily large enough. Thus, the  approximation for impulsive accelerated mirrors consists in setting $u_0\ll1$ and $k\gg1$, leading to the single main assumption  $ku_0=\text{const}$. The value of this constant is related to the final speed of the mirror itself. Indeed, the parameter $\nu$ in Eqs.~\eqref{accrbc} and Eq.~\eqref{acclbg} is simply given by  $\nu=e^{ku_0}$ for the Carlitz-Willey trajectory. The trajectories of such mirrors are portrayed in Fig. \ref{penrose}.
It is worth noticing that the above described approximation is valid for all those trajectories provided that we can associate to the acceleration of the mirror a constant parameter $k$.

\indent Finally we can apply the approximation to  Eqs.~\eqref{accrbc} and ~\eqref{acclbg}, using $k=\ln(\nu)/u_0$ and expanding in series for $u_0\rightarrow0$. For short acceleration periods, only the lowest expansion orders are clearly needful.

To the zeroth order the integrals on Eq.~\eqref{accrbc} and Eq.~\eqref{acclbg} can be completely neglected. In this case, the Bogoliubov coefficients read
\begin{align}
\label{0termR}
\beta_{\omega\omega'}^{RR}&=-\frac{\eta}{2\pi}\sqrt{\omega'\omega}\frac{\sqrt{\nu}\left(\omega'-\omega\sqrt{\nu}\right)\left(\omega\sqrt{\nu}-1\right)+i\eta\left(\nu-1\right)}{\left(\omega\sqrt{\nu}-i\eta\right)\left(\omega'+i\eta\right)\left(\omega\nu+\omega'\right)\left(\omega+\omega'\right)},\\
\beta_{\omega\omega'}^{RL}&=-\frac{\eta}{2\pi}\sqrt{\omega'\omega}\frac{1-\sqrt{\nu}}{\left(\omega\sqrt{\nu}-i\eta\right)\left(\omega'+i\eta\right)\left(\omega+\omega'\right)}.
\label{0termL}
\end{align}
Here, we are  neglecting the acceleration period and Eqs.~\eqref{0termR} and~\eqref{0termL} are valid for each trajectory for which we  associate a constant parameter $k$ to the acceleration. Both the Bogoliubov coefficients Eqs.~\eqref{0termR} and~\eqref{0termL} have the factor $\sqrt{\omega'\omega}$ at the beginning.

So, their modulus squares  show the factor $\omega'\omega$, that is not cancelled by any term in the denominator. This ensures that in these modulus squares no infrared divergences occur\footnote{This is valid for $\beta_{\omega\omega'}^{RR}$, if $\omega\ne0$. If not,  we could have possible  infrared divergences, leading to unphysical particle production. } for $\omega'\rightarrow0$. Furthermore, for $\omega'\rightarrow+\infty$ Eqs.~\eqref{0termR} and~\eqref{0termL} are asymptotic to $\omega'^{-3/2}$. This means that their modulus squares do not provide any ultraviolet divergence. As a consequence the number of particles produced with this approximation is finite and  different from zero for finite values of $\omega$. Finally, checking the case in which the mirror lies at rest, namely $\nu=1$, we immediately notice that the Bogoliubov coefficients are zero, as expected.\\
\indent To the first order expansion, we restrict the trajectories to the ones with non singular $\partial_u\overline{u}(u)$ and $\partial_u\overline{v}(v')$ throughout the range $0\le u\le u_0$. Hence, first order Bogoliubov coefficients are
\begin{widetext}
\begin{equation}\label{1orderR}
\beta_{\omega\omega'}^{RR}\sim\nonumber\\
-\frac{\eta}{2\pi}\sqrt{\omega'\omega}\left\{\frac{\sqrt{\nu}\left(\omega'-\omega\sqrt{\nu}\right)\left(\sqrt{\nu}-1\right)+i\eta\left(\nu-1\right)}{\left(\omega\sqrt{\nu}-i\eta\right)\left(\omega'+i\eta\right)\left(\omega\nu+\omega'\right)\left(\omega+\omega'\right)}+\frac{(\omega+\omega')\left[\eta\left(V_0\nu-1\right)+i\omega\nu\left(U_0\sqrt{\nu}-1\right)-i\omega'\sqrt{\nu}\left(V_0\sqrt{\nu}-U_0\right)\right]}{\left(\omega\sqrt{\nu}-i\eta\right)\left(\omega'+i\eta\right)\left(\omega\nu+\omega'\right)\left(\omega+\omega'\right)}u_0\right\},
\end{equation}
\end{widetext}
\begin{equation}\label{1orderL}
\beta_{\omega\omega'}^{RL}\sim-\frac{\eta}{2\pi}\sqrt{\omega'\omega}\frac{1-\sqrt{\nu}+i(\omega+\omega')\left(U_0\sqrt{\nu}-1\right)u_0}{\left(\omega\sqrt{\nu}-i\eta\right)\left(\omega'+i\eta\right)\left(\omega+\omega'\right)},
\end{equation}
where $U_0\coloneqq\partial_{u_0}\overline{u}_0|_{u_0=0}$ and $V_0\coloneqq\partial_{u_0}v_0|_{u_0=0}$. For the Carlitz Willey trajectory \cite{carlitz1987reflections}, we get
\begin{equation}
U_0=\frac{2}{\ln\nu}\left(1-\frac{1}{\sqrt{\nu}}\right),
\end{equation}
\begin{equation}
V_0=\frac{1}{\ln\nu}\left(1-\frac{1}{\nu}\right).
\end{equation}
In terms of $\nu$, the first order Bogoliubov coefficients for the Carlitz-Willey trajectory become
\begin{widetext}
\begin{equation}\label{betaCWR}
\begin{split}
\beta_{\omega\omega'}^{RR}\sim-\frac{\eta}{2\pi}\sqrt{\omega'\omega}\left\{\frac{\sqrt{\nu}\left(\omega'-\omega\sqrt{\nu}\right)\left(\sqrt{\nu}-1\right)+i\eta\left(\nu-1\right)}{\left(\omega\sqrt{\nu}-i\eta\right)\left(\omega'+i\eta\right)\left(\omega\nu+\omega'\right)\left(\omega+\omega'\right)}\right.\\
\left.+\frac{(\omega+\omega')\left[\eta\left(\frac{1}{\ln(\nu)}(\nu-1)-1\right)+i\omega\nu\left(\frac{2}{\ln(\nu)}\left(\sqrt{\nu}-1\right)-1\right)-i\omega'\frac{\sqrt{\nu}}{\ln(\nu)}\left(\frac{1}{\sqrt{\nu}}+\sqrt{\nu}-2\right)\right]}{\left(\omega\sqrt{\nu}-i\eta\right)\left(\omega'+i\eta\right)\left(\omega\nu+\omega'\right)\left(\omega+\omega'\right)}u_0\right\},
\end{split}
\end{equation}
\end{widetext}

\noindent and

\begin{align}\label{betaCWL}
&\beta_{\omega\omega'}^{RL}\sim\nonumber\\
&-\frac{\eta}{2\pi}\sqrt{\omega'\omega}\frac{1-\sqrt{\nu}+i(\omega+\omega')\left(\frac{2}{\ln(\nu)}\left(\sqrt{\nu}-1\right)-1\right)u_0}{\left(\omega\sqrt{\nu}-i\eta\right)\left(\omega'+i\eta\right)\left(\omega+\omega'\right)}.
\end{align}
Since only the first order of $u_0$ is considered, in the computation of the modulus square of the Bogoliubov coefficients, the terms proportional to $u_0^2$ are neglected.

\begin{figure*}
\centering
\subfigure[Spectrum of the particles produced by the mirror with $\eta=1$ for different values of $A$,  obtained integrating in $d\omega'$ Eq. \eqref{partprodintegrand}. It is $u_0=0.0001$.\label{tot}]{\includegraphics[height=0.3\hsize,clip]{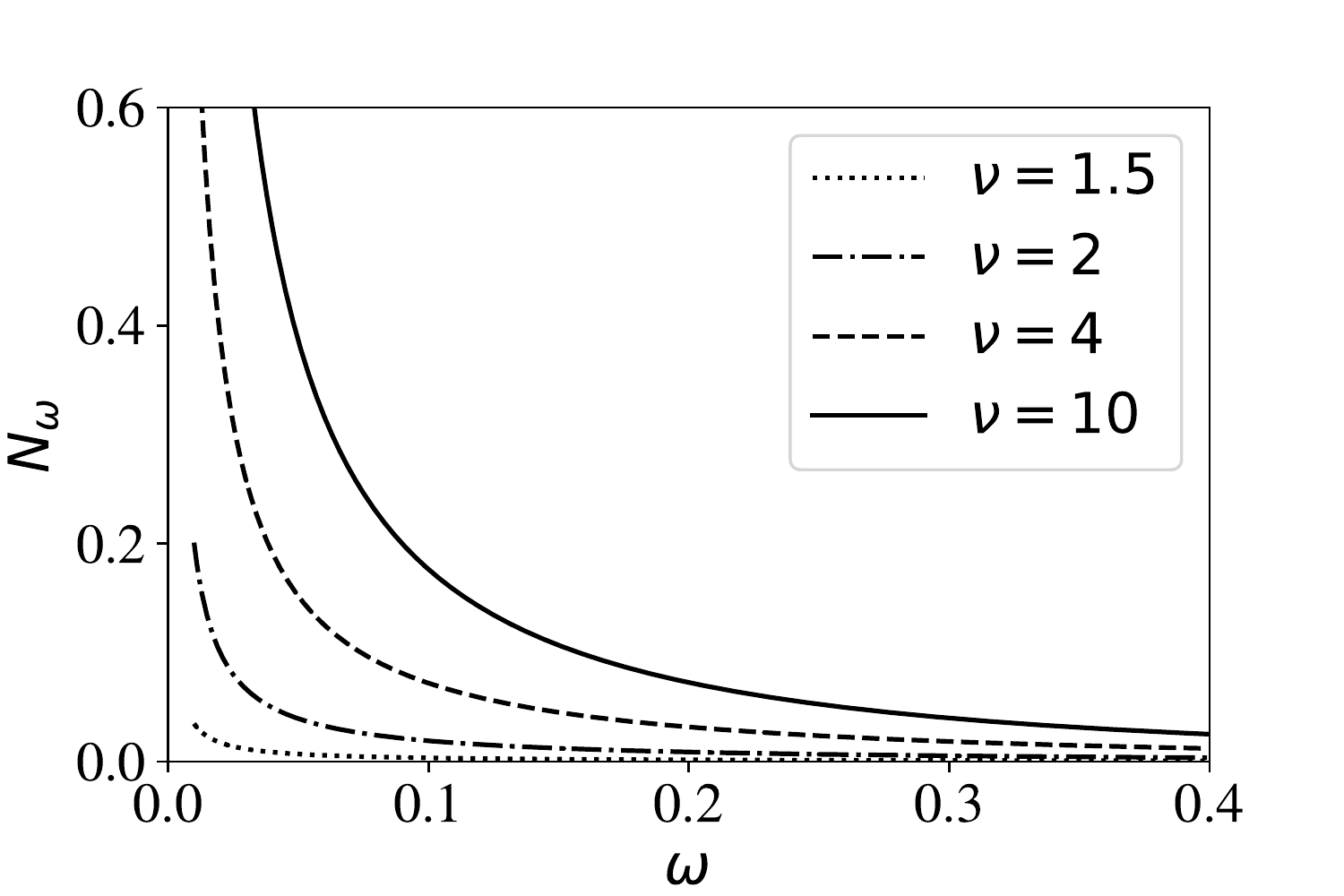}}
\subfigure[Comparison of the spectra of the particles produced at the right hand side of the mirror. $N_\omega^R$ (resp. $N_\omega^L$) is the contribution from the vacuum modes at the right (resp. left) hand side of the mirror, obtained integrating in $d\omega'$ the modulus squared of Eq.~\eqref{betaCWR} (resp. Eq.~\eqref{betaCWL}). The complete spectrum is the sum $N^L_\omega + N^R_\omega$. Here $\eta=1$, $u_0=0.0001$ and $A=0.1$.\label{comp1}]{\includegraphics[height=0.3\hsize,clip]{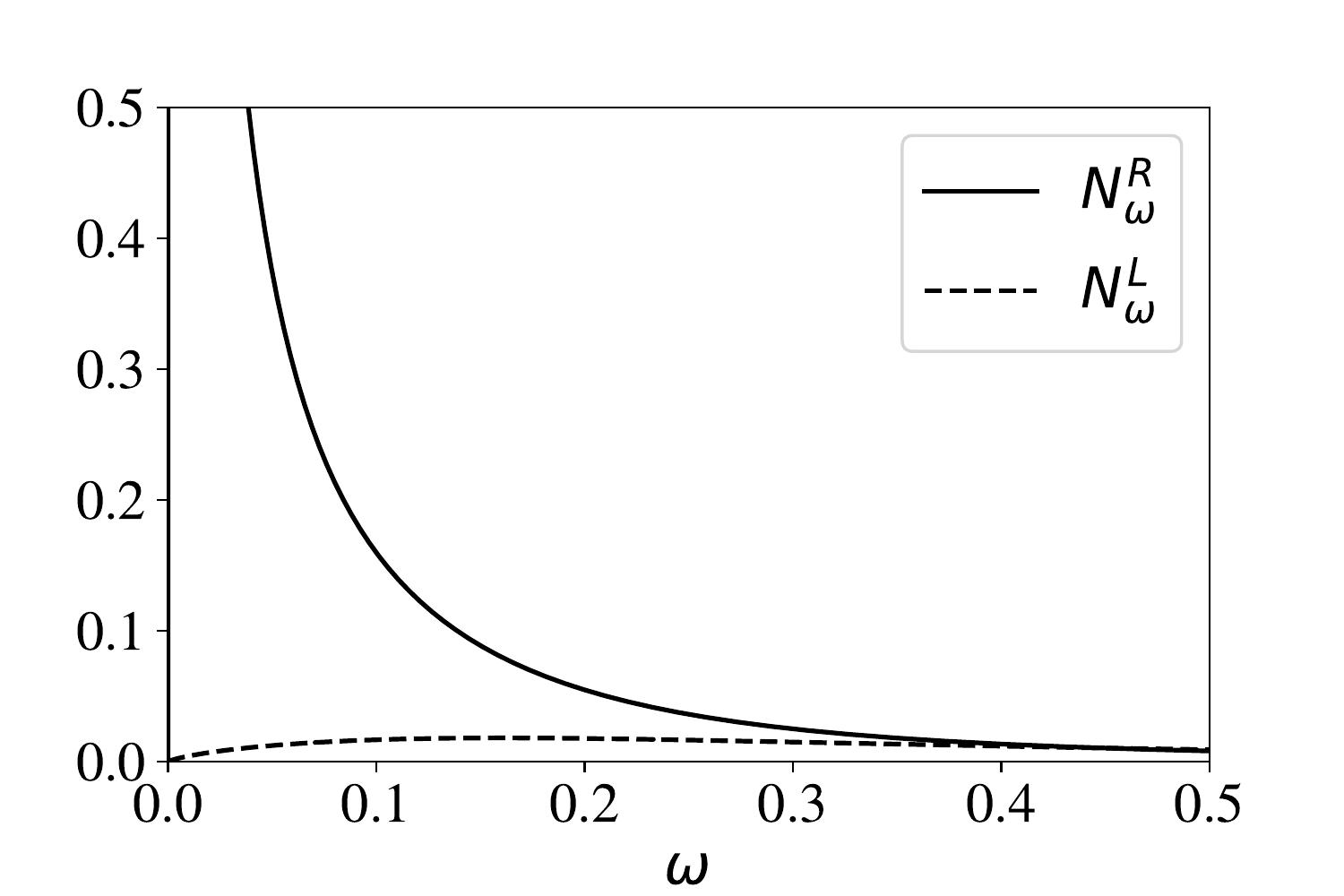}}
\subfigure[The same of \ref{comp1} but with $\eta=0.1$ and $A=0.5$.\label{comp2}]{\includegraphics[height=0.3\hsize,clip]{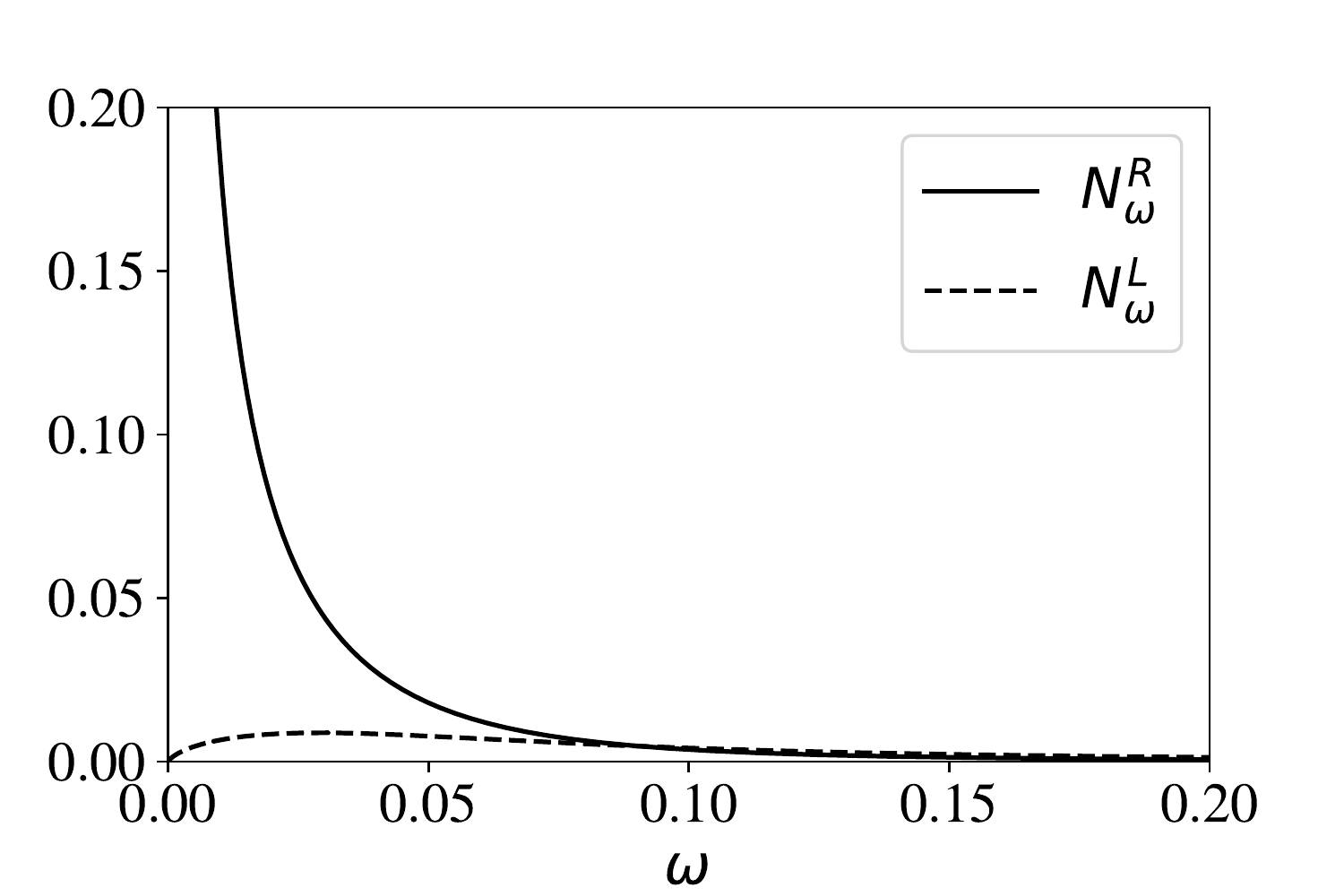}}
\subfigure[The same of \ref{comp1} but with $\eta=0.1$ and $A=0.1$.\label{comp3}]{\includegraphics[height=0.3\hsize,clip]{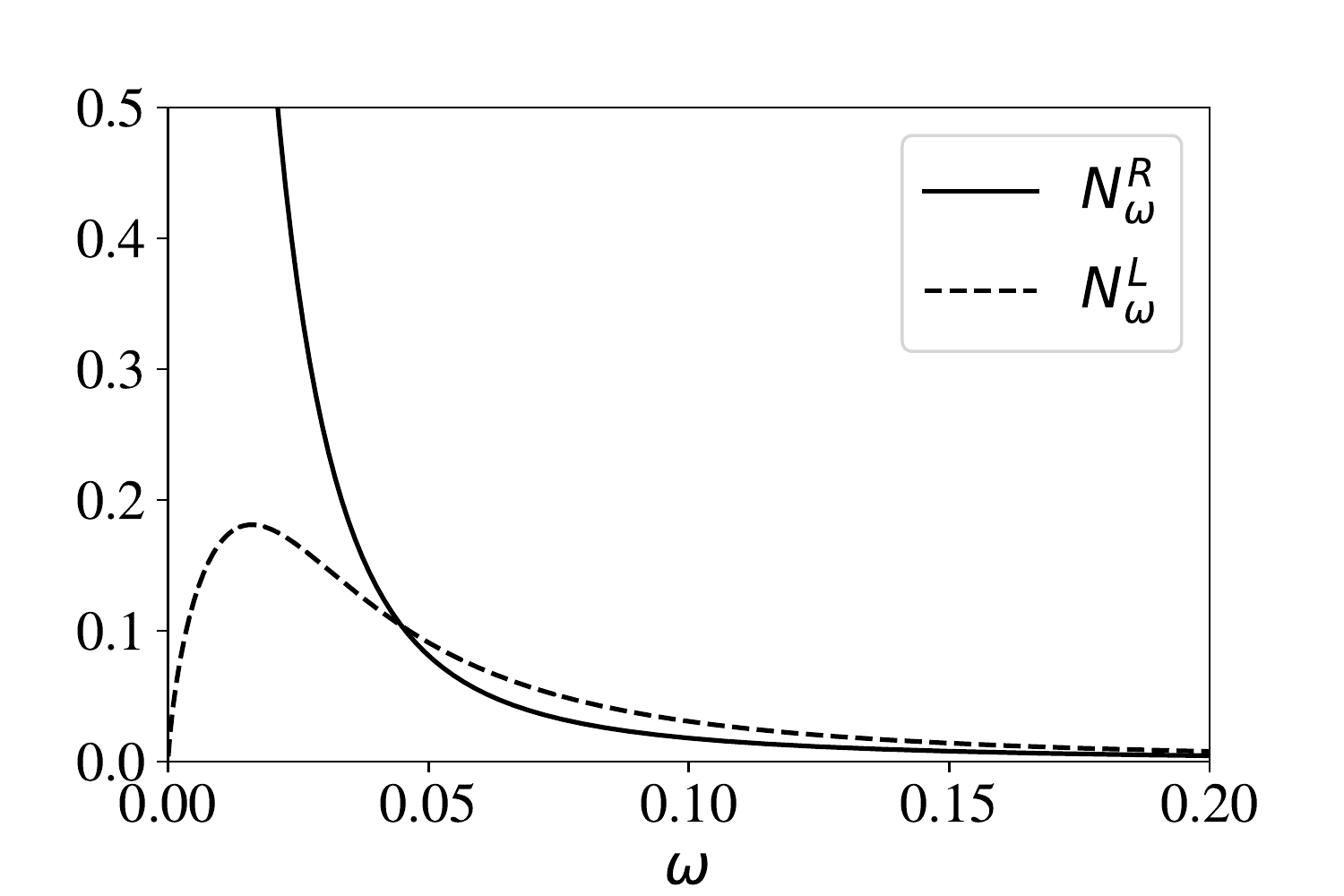}}
\caption{Plots showing the behaviour of the particle production as function of the frequency for a mirror with a trajectory like the ones shown in Fig.~\ref{penrose}.}
\label{figura}
\end{figure*}

\indent Summing the modulus square of Eq.~\eqref{betaCWL} and Eq.~\eqref{betaCWR}, we obtain
\begin{widetext}

\begin{equation}\label{partprodintegrand}
\begin{split}
\left|\beta_{\omega\omega'}^{RR}\right|^2+\left|\beta_{\omega\omega'}^{RL}\right|^2=\frac{\eta^2}{4\pi^2}\omega\omega'\left[\frac{(\sqrt{\nu}-1)^2(\omega\nu+\omega')^2+\nu(\omega\sqrt{\nu}-\omega')^2(\sqrt{\nu}-1)^2+\eta^2(\nu-1)^2}{(\omega^2\nu+\eta^2)(\omega'^2+\eta^2)(\omega\nu+\omega')^2(\omega+\omega')^2}\right.\\+\left.
\frac{2(\omega+\omega')\left(\eta\sqrt{\nu}(\omega'-\omega\sqrt{\nu})(\sqrt{\nu}-1)\left(\frac{1}{\ln(\nu)}(\nu-1)-1\right)+\eta(\nu-1)\left(\frac{\omega'}{\ln\nu}(\sqrt{\nu}-1)^2\right)-\omega\nu\left(\frac{2}{\ln\nu}(\sqrt{\nu}-1)-1\right)\right)}{(\omega^2\nu+\eta^2)(\omega'^2+\eta^2)(\omega\nu+\omega')^2(\omega+\omega')^2}u_0
\right].
\end{split}
\end{equation}

\end{widetext}
Thus, by integrating over $d\omega'$ we get the total number of particles with frequency $\omega$ created by the mirror. Even though the corresponding number could be analytically computed, its expression turns out to be extremely complicated. Thus, we omitted it explicitly.\\
Studying Eq.~\eqref{partprodintegrand}, one can prove that the zero order term increases as $\nu$ increases. In particular, it converges to an asymptotic value $\mathcal{A}$ as
$\propto\sqrt{\nu}^{\,-1}$. Instead, the first order term, as $\nu$ increases, goes to zero faster, namely as $\frac{1}{\sqrt{\nu}\ln{\nu}}$. Starting from this fact, one can prove that, for $u_0$ enough small, an upper bound for the particle production is provided when $\nu\to\infty$. This upper bound is provided by the following analytical expression.
\begin{equation}
N_\omega\rightarrow\frac{\eta^2}{2\pi^2\omega}\left[\frac{\omega^2-\eta^2}{(\omega^2+\eta^2)^2}\ln\left(\frac{\omega}{\eta}\right)+\frac{\pi\eta\omega-\omega^2-\eta^2}{(\omega^2+\eta^2)^2}\right].
\end{equation}
The spectrum of the particles production for the trajectories shown in Fig.~\ref{penrose} is depicted in Fig.~\ref{tot}. In the Figs.~\ref{comp1},~\ref{comp2},~\ref{comp3} some comparisons among the contributions of the right and left part of the mirror are shown: it can be seen from them that the two contributions are comparable when $\eta\ll u_0^{-1}$. We also stress an infrared divergence for the spectrum, in agreement with the modified Carlitz-Willey trajectory as found in Refs. \cite{B} and \cite{C}. Moreover, we can observe that the contribution of the vacuum modes in the right of the mirror dominates over the one in the left for low frequencies. For high frequencies the contribution of the vacuum modes in the left is slightly higher than the one in the right, becoming the same for $\omega\rightarrow\infty$ only.
\color{black}

\section{Quantum channel arising from a moving mirror}\label{Sec4}

We now revisit the model of semitransparent moving mirrors through an information communication perspective. In particular, the aim is to realize the field coming from left-past spacetime region, passing through the semitransparent moving mirror and ending up to right-future spacetime region as undergoing the action of a quantum channel. Then study the capacities in transmitting classical and quantum information of such a channel.

We consider as input a mode from $\mathcal{J}_L^-$ and as output the same mode transmitted by the mirror and outgoing toward $\mathcal{J}_R^+$. The bosonic modes in these two zones are related to each other through the Bogoliubov transformations, which are linear. Hence, taking the mode say of frequency $\omega$ as an input mode, while the other modes as environment modes (initially in vacuum, and after the process traced out), we end up with a Gaussian quantum channel.

To formalize such a mapping,  following  \cite{Giulio}, it is sufficient to know that a Gaussian quantum channel maps a bosonic Gaussian state into another bosonic Gaussian state. Considering only one mode $\omega$, the canonical variables in phase space are given by the vector $\mathbf{z}_\omega=(q_\omega,p_\omega)\in\mathbb{R}^2$ and a Gaussian bosonic state is represented by a characteristic function
\begin{equation}\label{carbeforechannel}
    \chi^{in}({\bf z}_\omega)=\exp\left(-\frac{1}{4}{\bf z}_\omega^T\cdot\sigma_\omega\cdot{\bf z}_\omega+i{\bf d}_\omega^T\cdot{\bf z}_\omega\right),
\end{equation}
where ${\bf d}_\omega=\left(\langle Q_\omega\rangle,\langle P_\omega\rangle\right)$ and $\sigma_j$ is the so called covariance matrix, defined as\footnote{$Q_\omega=\frac{1}{\sqrt{2}}\left(a_\omega^\dagger+a_\omega\right)$ and $P_\omega=\frac{1}{i\sqrt{2}}\left(a_\omega-a_\omega^\dagger\right)$ are the canonical quadrature operators for the mode $\omega$.}
\begin{equation}
\sigma_\omega=\left(\begin{matrix}
\langle Q_\omega^2\rangle & \frac{1}{2}\langle P_\omega Q_\omega+Q_\omega P_\omega\rangle\\\frac{1}{2}\langle P_\omega Q_\omega+Q_\omega P_\omega\rangle& \langle P_\omega^2\rangle
\end{matrix}\right).\label{comatrix}
\end{equation}

A one mode Gaussian quantum channel maps the one mode characteristic function, Eq.~\eqref{carbeforechannel}, into
\begin{equation}\label{carafterchannel}
\begin{split}
    \chi^{out}({\bf z}_\omega)=\exp\big(-\frac{1}{4}\mathbf{z}_\omega^T\cdot\left(\mathbb{T}\sigma_\omega\mathbb{T}^T+\mathbb{N}\right)\cdot\mathbf{z}_\omega\\
    +i\left(\mathbf{d}_\omega^T\cdot\mathbb{T}^T+\mathbf{v}_\omega^T\right)\cdot\mathbf{z}_\omega\big),
\end{split}
\end{equation}
where a Gaussian quantum channel is so characterized by the triad $\left(\mathbb{T}, \mathbb{N}, \mathbf{v}\right)$ with $\mathbb{T}$ and $\mathbb{N}$ as two $2\times2$ matrices respectively related to the attenuation/amplification of the mode, and to the noise which affects the input signal.  In particular, the attenuation/amplification is given by $\tau=\det\mathbb{T}$. Since the information transmission capabilities of the channel can be characterized in terms of entropic quantities, that do not depend on the vector $\mathbf{v}$, we can investigate the evolution of the covariance matrix, $\sigma_j$. Thus,  from Eqs.~\eqref{carbeforechannel} and~\eqref{carafterchannel}, we write
\begin{equation}\label{covmap}
 \sigma_\omega^{in}\longmapsto\sigma_\omega^{out}=\mathbb{T}\sigma_\omega^{in}\mathbb{T}^T+\mathbb{N}.
\end{equation}
Our focus is on the covariance matrix, Eq.~(\ref{comatrix}), of the input mode and its output as a result of Eq.~(\ref{covmap}). In line with our aim at the beginning of this section, all modes at $\mathcal{J}_L^-$ are considered uncorrelated and in the vacuum, but the single frequency mode of interest is $\omega$. This can be formalized with the following values
\begin{equation}\label{1imp}
\langle a_{\omega'}^La_{\omega''}^L\rangle=M\delta(\omega-\omega')\delta(\omega'-\omega''),
\end{equation}
\begin{equation}\label{2imp}
\langle a_{\omega'}^{L\dagger}a_{\omega''}^L\rangle=N\delta(\omega-\omega')\delta(\omega'-\omega''),
\end{equation}
and the expectation values of all the other combination of bosonic operators (including the ones relative to the right side of the mirror) equal to zero (except for the Hermitian of Eq.~\eqref{1imp} and the commutation of Eq.~\eqref{2imp}). Moreover, following Eqs.~\eqref{1imp} and \eqref{2imp}, $N$ is the mean number of particles in the input state, and $M$ specifies the correlation between $Q_\omega$ and $P_\omega$. \\
\indent Using Eqs.~(\ref{1imp}) and (\ref{2imp}), we calculate $\sigma_\omega^{in}$ and $\sigma_\omega^{out}$ taking the input and output bosonic operator for the mode $\omega$ (related to each other by a Bogoliubov transformation). We calculate the quadrature operators for both of them using Eq.~\eqref{1imp} and Eq.~\eqref{2imp}. It turns out that the input and output covariance matrices are related by a relation equal to Eq.~\eqref{covmap} from which we can obtain the entries of the $\mathbb{T}=\left(\begin{matrix}T_1&T_2\\T_3&T_4
\end{matrix}\right)$ and $\mathbb{N}=\left(\begin{matrix}N_1&N_2\\N_3&N_4
\end{matrix}\right)$ in terms of Bogoliubov coefficients, i.e.,
\begin{align}
&T_1=\epsilon\pi\Re\left(\alpha_{\omega\omega}^{RL}-\beta_{\omega\omega}^{RL}\right)\,,\\
&T_2=\epsilon\pi\Im\left(\alpha_{\omega\omega}^{RL}+\beta_{\omega\omega}^{RL}\right)\,,\\
&T_3=-\epsilon\pi\Im\left(\alpha_{\omega\omega}^{RL}-\beta_{\omega\omega}^{RL}\right)\,,\\
&T_4=\epsilon\pi\Re\left(\alpha_{\omega\omega}^{RL}+\beta_{\omega\omega}^{RL}\right)\,,
\end{align}

\begin{align}\label{Ns}
N_1=&-\frac{\epsilon\pi}{2}\left|\alpha_{\omega\omega}^{RL*}-\beta_{\omega\omega}^{RL}\right|^2+\\
&+\frac{1}{2}\int_{0}^{\infty}\left(\left|\alpha_{\omega\omega'}^{RL*}-\beta_{\omega\omega'}^{RL}\right|^2+\left|\alpha_{\omega\omega'}^{RR*}-\beta_{\omega\omega'}^{RR}\right|^2\right)d\omega'\,,\nonumber\\
\label{Ns2}N_2=&N_3=-\epsilon\pi\Im\left(\alpha_{\omega\omega}^{RL}\beta_{\omega\omega}^{RL}\right)+\\
&+\int_{0}^{\infty}\Im(\alpha_{\omega\omega'}^{RL}\beta_{\omega\omega'}^{RL}+\alpha_{\omega\omega'}^{RR}\beta_{\omega\omega'}^{RR})d\omega'\,,\nonumber\end{align}
\begin{align}\label{Ns3}
N_4=&-\frac{\epsilon\pi}{2}\left|\alpha_{\omega\omega}^{RL*}+\beta_{\omega\omega}^{RL}\right|^2+\\
&+\frac{1}{2}\int_{0}^{\infty}\left(\left|\alpha_{\omega\omega'}^{RL*}+\beta_{\omega\omega'}^{RL}\right|^2+\left|\alpha_{\omega\omega'}^{RR*}+\beta_{\omega\omega'}^{RR}\right|^2\right)d\omega',\nonumber
\end{align}
where $\epsilon$ is the cutoff\footnote{We ignored it for  $\beta$  coefficients, although it is needful for  $\alpha$ coefficients, otherwise as $\omega=\omega'$ possible divergences arise. } of Eqs.~\eqref{accrbc} and~\eqref{acclbg}. Here $\Re$ and $\Im$ denote the real and imaginary parts, respectively. \\
\indent For the average attenuation/amplification in time $\tau\coloneqq\det\mathbb{T}$, we have the following general expression
\begin{equation}\label{taugeneric}
    \tau=\epsilon^2\pi^2\left(|\alpha_{\omega\omega}^{RL}|^2-|\beta_{\omega\omega}^{RL}|^2\right).
\end{equation}
Applying the Bogoliubov coefficients of an impulsive accelerated mirror to first order in a $u_0$ expansion around $u_0=0$ and using the Carlitz-Willey's acceleration we get
\begin{equation}\label{tau}
\tau=\frac{4\Omega^4\nu+\Omega^2\left(1+\sqrt{\nu}\right)^2}{4\left(\Omega^2\nu+1\right)\left(\Omega^2+1\right)},
\end{equation}

where $\Omega\coloneqq\omega/\eta$. It is easy to see that $\tau = \frac{\omega^2}{\omega
^2+\eta
^2}$,  as expected, for $\nu=1$. We are reminded that as $\Omega\to \infty$ one has $\tau =1$ (perfect transparency) and as $\Omega \to 0$, one has $\tau\to 0$ (perfect reflection), even when $\nu\neq 1$.

Transmission through a semitransparent mirror should result in three main effects:
\begin{itemize}
    \item[{\bf 1.}] \emph{a loss of the input signal, since part of that is reflected};
    \item[{\bf 2.}] \emph{an interference of the input signal with the other modes of the initial vacuum environment};
    \item[{\bf 3.}] \emph{a particle production contribution which eventually amplifies the input signal}.
\end{itemize}
One can prove that $\tau<1$ for all frequencies and $\nu$, providing that no signal amplification occurs, as evidence of the first effect. Nevertheless, for $\nu>1$, there is a reduction of such loss. The second and third effect also show themselves in this way (nevertheless, the effect of the particle production mostly arises as noise, since it occurs even without an input signal).

Another relevant aspect of Eq. \eqref{tau} is that to the first order in $u_0$ around $u_0=0$ does not contribute to $\tau$. In general, one can prove that also the other orders of $u_0$ does not give a contribution to $\tau$ in the limit $\epsilon\rightarrow0$. In fact, calculating $\tau$ from the general expression Eq.~\eqref{acclbg} (taking the corresponding $\alpha$ coefficient), for finite acceleration periods result in convergent integrals. Under these conditions, in the limit $\epsilon\rightarrow0$, only the divergent part of $\alpha_{\omega\omega}^{RL}$ gives a contribution:
\begin{equation}
\lim_{\epsilon\rightarrow\infty}\alpha_{\omega\omega}^{RL}=\frac{1}{\pi\epsilon}-\frac{\eta}{2\pi\epsilon}\left[\frac{\left(1+\sqrt{\nu}\right)\omega-2i\eta}{i\left(\omega\sqrt{\nu}-i\eta\right)\left(\omega-i\eta\right)}\right].
\end{equation}
After a calculation, one can prove that this will lead to the same $\tau$ of Eq.~\eqref{tau}. In conclusion, we have shown that, for the calculation of $\tau$ we are justified in removing the restriction over a small acceleration period.  It is then realized that this $\tau$
is valid for any mirror trajectory with a finite acceleration period, at least in the limit $\epsilon\rightarrow0$.\\

\begin{figure}
    \centering
    \includegraphics[scale=0.6]{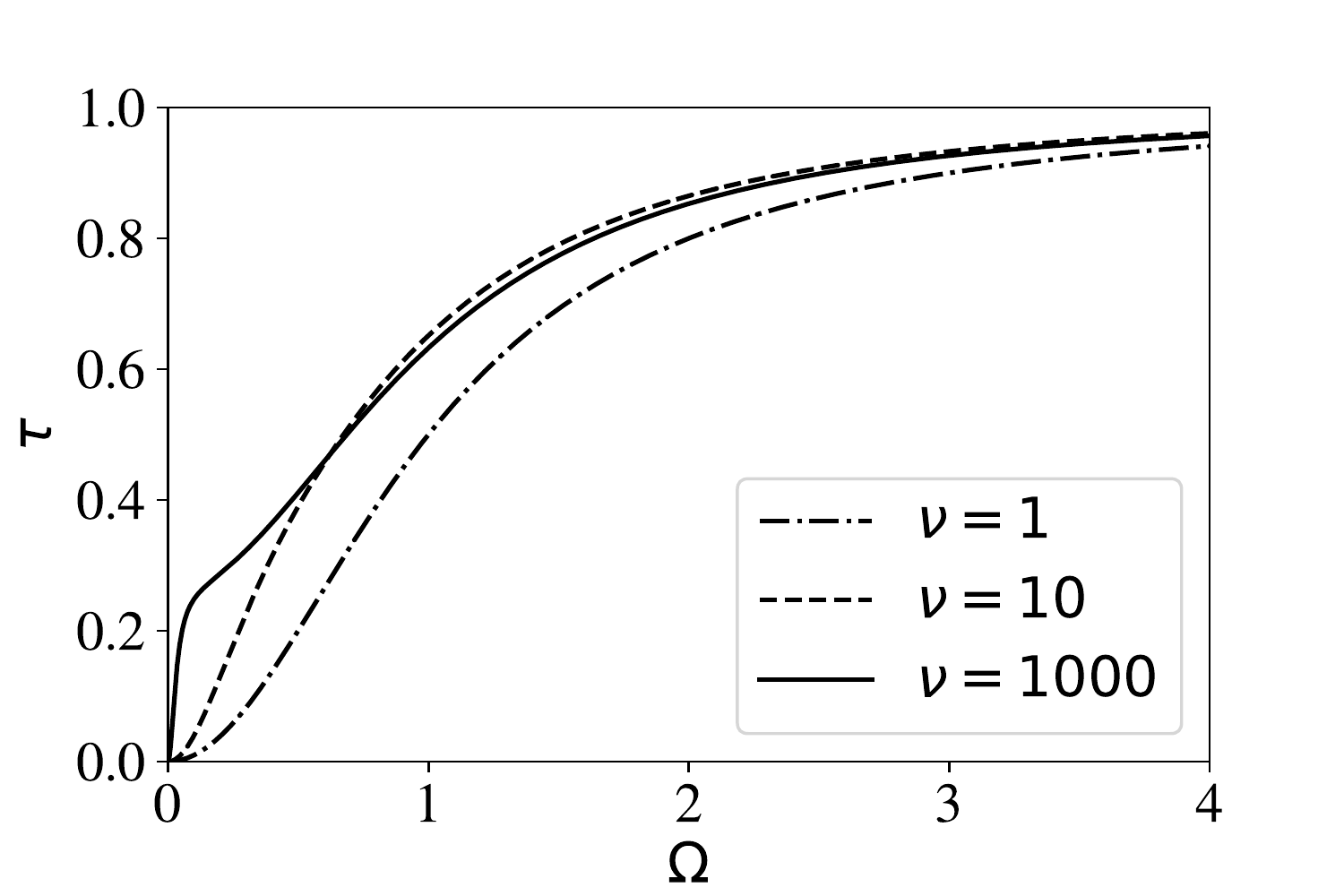}
    \caption{Behaviour of $\tau$ vs $\Omega$ from Eq.~\eqref{tau}. It was considered: a static mirror ($\nu=1$), a mirror accelerating toward the left with a final speed comparable to $V_{\textrm{crit}}\sim 0.8$, i.e. the critical speed at high frequencies ($\nu=10$) and a mirror accelerating toward the left with a final speed really close to the speed of light ($\nu=1000$).}
    \label{tau figure}
\end{figure}

Studying $\tau$ from Eq. \eqref{tau}, for $\nu\rightarrow\infty$, we have an asymptotic behaviour of $\tau$ equal to
\begin{equation}\label{asymptotictau}
\tau=\frac{\Omega^2+\frac{1}{4}}{\Omega^2+1}.
\end{equation}
However, differently from the particle production $N_\omega$ (see Sec.~\ref{Sec3}), $\tau$ does not increase in a monotonic way by increasing the final speed of the mirror toward its left (as considered in Fig. \ref{penrose}). In fact, for each value of $\Omega$ we have a finite value of $\nu$ which maximizes $\tau$. In other words, it exists a critical final $V_{\textrm{crit}}(\Omega)$ of a mirror (accelerating toward the left) with finite acceleration period, for which $\tau$ reaches a maximum and slightly decreases for $V>V_{\textrm{crit}}$, asymptotically reaching  $\tau$ described by Eq. \eqref{asymptotictau}. The critical value of $\nu$, say $\nu_{\textrm{crit}}$, in function of $\Omega$, can be written as
\begin{equation}\label{critA}
\nu_{\textrm{crit}}(\Omega)=\frac{9}{2}+\frac{4}{\Omega^2} +\frac{1+(3\Omega^2+1)\sqrt{(9\Omega^2+1)(\Omega^2+1)}}{2\Omega^4}.
\end{equation}
For $\Omega\ll1$ we have $\nu_{\textrm{crit}}\rightarrow\infty$ which corresponds to $V_{\textrm{crit}}=1$. For $\Omega\gg1$ the critical speed reaches the asymptotic value $V_{\textrm{crit}}(\Omega\to\infty)=0.8$. The decreasing of $\tau$ after $V_{\textrm{crit}}$ is  sharper for low frequencies, although $V_{\textrm{crit}}$ is  closer to the speed of light in this range. A plot for $\tau$ is portrayed in Fig. \ref{tau figure}.

One expects a contribution to the noise created by the mirror due to the particle production and characterized by $N_\omega$ with frequency mode $\omega$. Since $\tau<1$ for each $\omega$, $\eta$ and $A$, we anticipate a lossy and noisy quantum channel. By the classification of one-mode Gaussian channels made in Ref. \cite{Holevo}, we perform two unitary Gaussian transformations, one before and one after the quantum channel (respectively named, pre-processing and post-processing), in order to reduce the matrix $\mathbb{N}$, of a lossy and noisy quantum channel, to its canonical form
\begin{equation}\label{Ncan}
\mathbb{N}_c=(1-\tau)\left(\frac{1}{2}+\overline{n}\right)\mathbb{I},
\end{equation}
where $\overline{n}$ is the number of noisy particles created by the quantum channel. The term ${1\over2}$ is naturally related to vacuum energy induced by mode $\omega$. In the continuous limit we expect that it takes the value $\frac{1}{2\pi\epsilon}$.

Moreover, instead of finding the average number of noisy particles arriving to the detector $\overline{n}$, in the continuous limit we expect to have a spectrum of particles  expressed as $\overline{n}_\omega$. The former, once  integrated in a range of frequencies, provides a dimensionless number. Hence, in the continuous case,  Eq.~\eqref{Ncan} becomes
\begin{equation}\label{contNcan}
\det\mathbb{N}=\left(1-\tau\right)\left(\frac{1}{2\pi\epsilon}+\overline{n}_\omega\right).
\end{equation}
In the continuous limit only an infinitesimal range of frequencies would be detected. We can therefore write $\overline{n}=\epsilon\pi\overline{n}_\omega$.\\
\indent Since the determinant of $\mathbb{N}$ does not change when reducing it to its canonical form, we can study it from  Eqs.~\eqref{Ns}, \eqref{Ns2} and \eqref{Ns3}, leading to
\begin{equation}\label{Ncont}
\det\mathbb{N}=\det\mathbb{N}_c=\left(\frac{1-\tau}{2\pi\epsilon}+B\right)^2-C^2,
\end{equation}
where

\begin{equation}
B\coloneqq-\epsilon\pi\left|\beta_{\omega\omega}^{RL}\right|^2+N_\omega,
\end{equation}
and
\begin{equation}
C\coloneqq\left|\epsilon\pi\alpha_{\omega\omega}^{RL}\beta_{\omega\omega}^{RL}-\sum_{S=L,R}\int_{0}^{\infty}\alpha_{\omega\omega'}^{RS}\beta_{\omega\omega'}^{RS}d\omega'\right|.
\end{equation}
Comparing Eq.~\eqref{contNcan} with Eq.~\eqref{Ncont}, we get
\begin{equation}\label{spectrumnoisy}
    \overline{n}_\omega=\frac{1}{2\pi\epsilon}\left[-1+\sqrt{1+4\pi\epsilon\frac{B}{1-\tau}+4\pi^2\epsilon^2\frac{B^2-C^2}{(1-\tau)^2}}\right].
\end{equation}
Thus, the average number of noisy particles arriving to the detector is
\begin{equation}\label{contlimitnoise}
    \overline{n}=\frac{1}{2}\left[-1+\sqrt{1+4\frac{\tilde{B}}{1-\tau}+4\frac{\tilde{B}^2-\tilde{C}^2}{(1-\tau)^2}}\right],
\end{equation}
where $\tilde{B}\coloneqq\epsilon\pi B$ and $\tilde{C}=\epsilon\pi C$.\\
\indent If $B$ and $C$ are not divergent, for $\epsilon\rightarrow0$ we have $B=N_\omega$ and expanding the square root in the last term of Eq.~\eqref{spectrumnoisy}, we get the spectrum
\begin{equation}\label{overn}
    \overline{n}_\omega=\frac{N_\omega}{1-\tau}.
\end{equation}
Consequently, in this case, $\overline{n}=0$, that corresponds to our stand-alone approximation, i.e., to an impulsive accelerated mirror. Indeed, we already demonstrated $N_\omega$ is convergent as well as $B$, with $\tilde{B}=0$. Analogously, one can easily prove the convergence of $C$ as well, leading to $\tilde{C}=0$. This behaviour naturally suggests that the impulsive semitransparent mirror acts like a beam splitter.

A different expectation  occurs when $B$ and $C$ are divergent, e.g. for the perfectly reflecting Carlitz-Willey mirror. In this case a rigorous approach to get $\tilde{B}$ and $\tilde{C}$ requires the use of wave packets, where the frequency range is supposed to vanish. In fact, applying this approach to the Carlitz-Willey trajectory furnishes a finite $\tilde{B}$, see e.g. \cite{B} for further details.


\section{Quantum channel capacities}\label{Sec5}

In this section, we evaluate  classical and quantum capacities of the quantum channel described in the previous section. In so doing, we quantify the capability of an impulsive accelerated mirror to transmit both classical and quantum information.

For bosonic Gaussian channels the regularization of the capacities is a hard task and this problem is not fully solved, neither for classical nor quantum capacities. Fortunately, the channel we obtained in Sec. \ref{Sec4} becomes a beam splitter in the continuous limit. For this kind of channels the additivity is proved both for  classical  \cite{LossyCapacity} and  quantum capacity \cite{Calculationoneshot}.

We start by studying the classical capacity. Let us take the classical information we want to transmit with continuous random variable $X$ and probability distribution $p_x$. The encoding procedure is identified by a map which associates to each value $x$ of the random variable a state $\rho_x$. Let  $\Phi$ be the quantum channel of communication. The maximum that we can extract about $X$ at the channel output is given by \emph{Holevo information} \cite{Holevoinf1,Holevoinf2}:
\begin{equation}
\chi(\rho,\Phi)=S\left(\Phi(\rho)\right)-\int p_xS(\Phi(\rho_x))dx.
\end{equation}
where $S$ is the von Neumann entropy and $\rho\coloneqq\int p_x\rho_xdx$.\\
For one-mode Gaussian (OMG) channels it is possible to express the Holevo information in terms of covariance matrices if we restrict the possible encodings to Gaussian ones, see e.g. \cite{OMGclassicalc}. Namely, we have to restrict the possible inputs of the OMG channel $(\mathbb{T},\mathbb{N},\mathbf{v})$ to be bosonic Gaussian states with covariance matrix $\sigma$ and $\mathbf{d}=(x,0)$. Moreover we assume $p_x$ to be a Gaussian probability distribution with mean equal to zero and covariance matrix $\sigma'$. In the reference \cite{ConjectureGaussian} it is proved that, if the channel is a beam splitter, such encodings maximize the Holevo information,  becoming \cite{HolevoCM}
\begin{equation}\label{Holevoinformation}
S\left(\mathbb{T}(\sigma+\sigma')\mathbb{T}^T+\mathbb{N}\right)-S\left(\mathbb{T}\sigma\mathbb{T}^T+\mathbb{N}\right),
\end{equation}
where the Von Neumann entropy $S$, referring to a covariance matrix $\sigma$, can be written by $S(\sigma)=h(d)$, with $d\coloneqq\sqrt{\det(\sigma)}$ and
\begin{equation*}
S(\sigma)=\left(d+\frac{1}{2}\right)\log\left(d+\frac{1}{2}\right)-\left(d-\frac{1}{2}\right)\log\left(d-\frac{1}{2}\right).
\end{equation*}
At this point, the classical capacity $C$ is given by the maximum of  Eq.~\eqref{Holevoinformation} over the inputs $\sigma$ and $\sigma'$. However, since the bosonic Gaussian states are in an infinite-dimensional Hilbert space, as input we can take a state with an infinite particle amount. Obviously, this maximizes the Holevo information and it leads to an infinite classical capacity. This case is unrealistic, since we need an infinite amount of energy for the encoding process. In order to remove this possibility, we have to impose a restriction on the maximum energy $E$ which can be used for the encoding, by
\begin{equation}
    \frac{1}{2}\omega\text{Tr}(\sigma+\sigma')\le E.
\end{equation}

\begin{figure}
    \centering
    \includegraphics[scale=0.6]{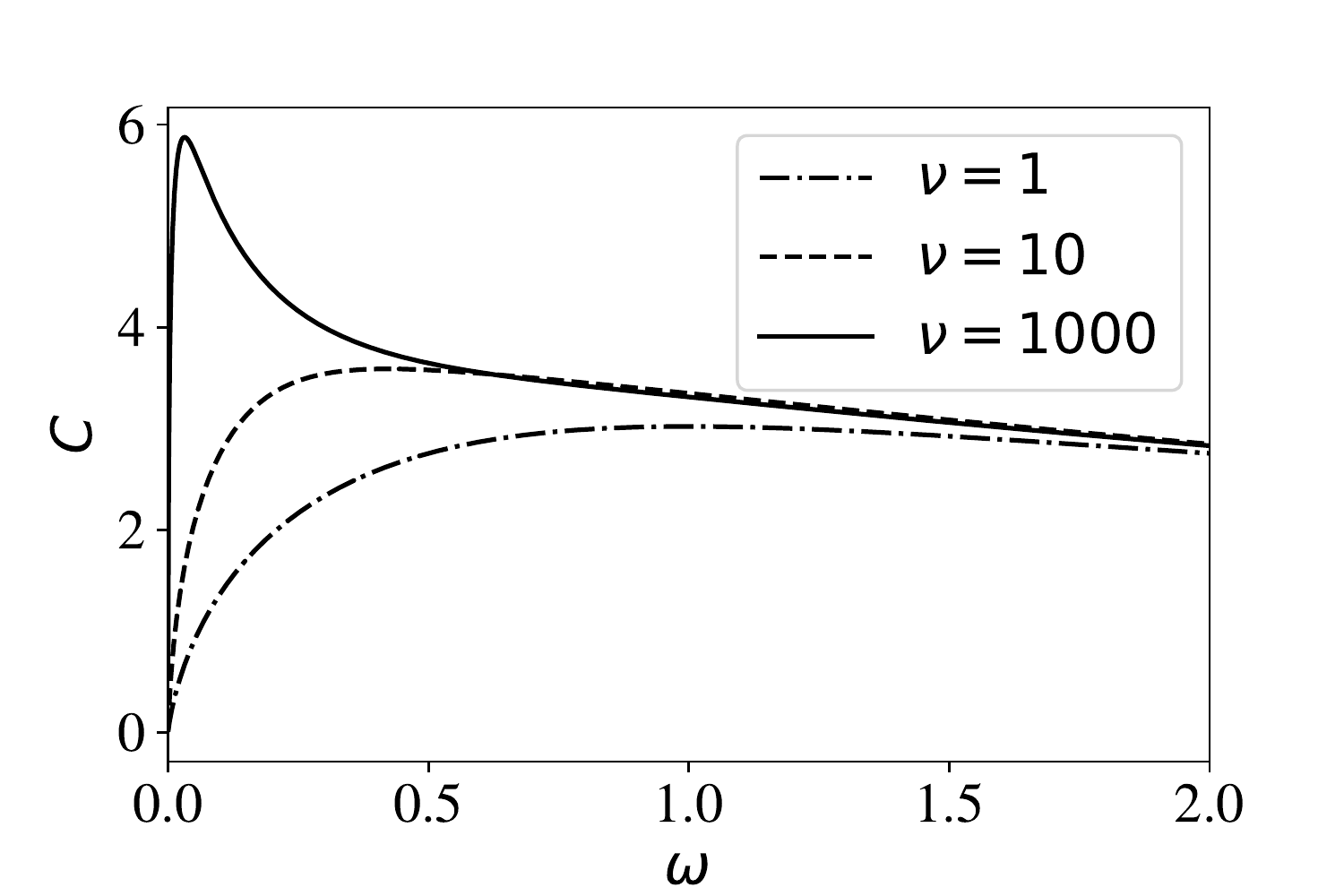}
    \caption{Classical capacity of an impulsive accelerated mirror constraining the encoding energy to $E=5$ according to Eq.~\eqref{cgused} for various values of $\nu$. $\tau$ is provided by the Eq.~\eqref{tau} with $\eta=1$. }
    \label{ClassicalCapacity}
\end{figure}

With this prescription, the following classical capacity for a noiseless, lossy channel (beam splitter), has been obtained by  \cite{LossyCapacity}
\begin{equation}\label{cgused}
\begin{split}
    C&=\frac{\tau E}{\omega}\log\left(\frac{\tau E+\omega}{\tau E}\right)+\log\left(\frac{\tau E+\omega}{\omega}\right).
\end{split}
\end{equation}
Plots of $C$ are shown is Fig. \ref{ClassicalCapacity}. Since the channel is asymptotically without loss for $\omega\rightarrow\infty$, one can expect a constant capacity in this limit. However, for great frequencies the encoding necessitates more energy. For this reason, if the energy is constrained, the number of photons we can use for the encoding  decreases linearly. In fact, studying the asymptotic behaviour of $\tau$ for $\omega\rightarrow\infty$, it turns out that $C$ goes to zero very slowly,  $\sim\frac{\ln\omega}{\omega}$. Instead, for $\omega\rightarrow0$, $C$ goes to zero linearly. Moreover, from Eq.~\eqref{cgused} we can see that the capacity increases with $\tau$.

As a consequence, we expect the capacity to be maximized for $\nu=\nu_{\textrm{crit}}$. Further, it is interesting to notice, from Fig. \ref{ClassicalCapacity}, how for different values of $\nu$ the maximum of the lower bound of the classical capacity occurs for different values of $\omega$. This ``maximum capacity frequency" seems to decrease when we increase $\nu$. However, since the peak becomes higher increasing the final speed of the mirror, we can conclude that the capability of an impulsive accelerated mirror to transmit classical information always increases with the mirror final speed.\\
Next, we proceed to evaluate the quantum capacity, obtained maximizing the coherent information over the input (for a beam-splitter channel, the quantum capacity is additive \cite{HolevoCM}). Br\'adler \cite{OMGquantumc} proved that the one-shot quantum capacity of a lossy and noisy OMG channel is maximized: either when the number of photons $N$ used for the encoding is null, or when this number $N$ is infinite (so, we have an infinite amount of energy for the encoding). In the first case the maximized coherent information is zero and no quantum information can be transmitted reliably. However, unlike the classical capacity framework, in case of infinite amount of energy of the encoding we have a finite value for the coherent information. This means that there is no need to impose a constraint for the energy of the encoding in order to have a finite value for the one-shot quantum capacity. Nevertheless, the infinite energy of the encoding is unrealistic. For this reason,  the quantum capacity that we intend to study might be considered as upper bound of the ``real one" with a finite encoding energy. However, the quantum capacity,  obtained with a finite $E$, is basically the same of the one obtained with $E\rightarrow\infty$ in the region $\omega\ll E$. Consequently, even if we compute a quantum capacity for $E\rightarrow\infty$, it is realistic to choose $E$ large enough within the range of employed frequencies.

\begin{figure}
    \centering
    \includegraphics[scale=0.6]{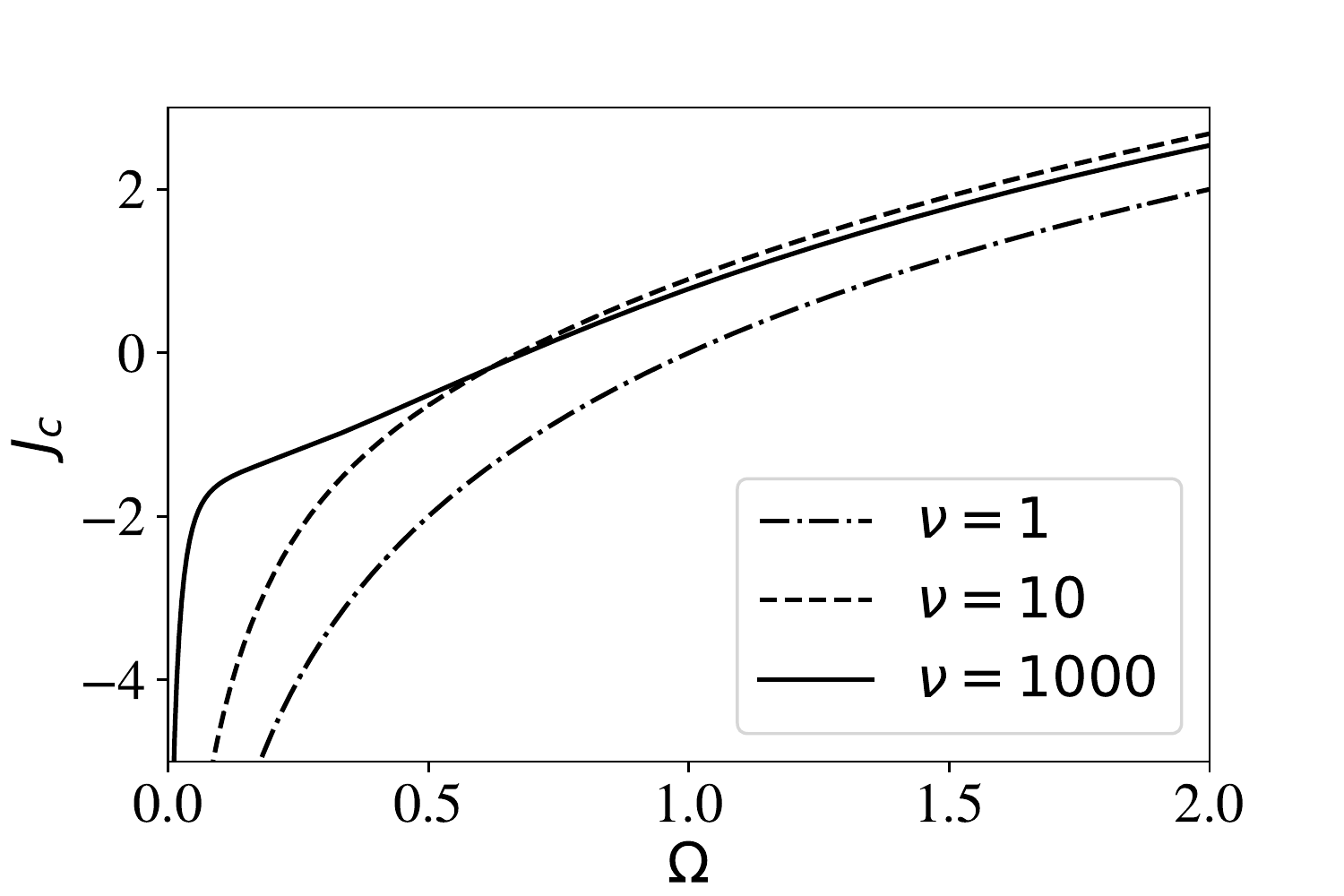}
    \caption{Coherent information for the impulsive accelerated mirror with an input having an infinite number of photons, according to Eq.~\eqref{coherentmirror}.}
    \label{QuantumCapacity}
\end{figure}

For the coherent information $J_c$ of a OMG lossy channel, as $E\rightarrow\infty$ we have  \cite{Holevo,Calculationoneshot}
\begin{equation}
J_c(E\rightarrow\infty)=\log\frac{\tau}{1-\tau}.
\end{equation}
Using $\tau$ from Eq. \eqref{tau} we get
\begin{equation}\label{coherentmirror}
    J_c(E\rightarrow\infty)=\log\left[\Omega^2\frac{\frac{(1+\sqrt{\nu})^2}{4}+\Omega^2\nu}{1+\Omega^2\left(\frac{3}{4}-2\sqrt{\nu}+\frac{3}{4}\nu\right)}\right],
\end{equation}
whose behaviour is shown in Fig.  \ref{QuantumCapacity}.
The quantum capacity of an impulsive accelerated mirror is given by
\begin{equation}\label{osqc}
Q(\omega)=\max\{0,J_c(E\rightarrow\infty,\omega)\}.
\end{equation}
Analyzing Eq.~\eqref{coherentmirror}, we have that $Q$ diverges logarithmically as $\omega\rightarrow\infty$, in agreement with the fact that the channel becomes without loss in this limit. Even in this case, for each frequency, the maximum of the quantum capacity is obtained for $\nu=\nu_{\textrm{crit}}$. The quantum capacity is non-zero only if $\tau>\frac{1}{2}$. This is in agreement with the no-cloning theorem, for which it is impossible to transmit quantum information reliably if we have a loss $1-\tau$ larger than $\frac{1}{2}$, otherwise the input state can be cloned, (see e.g. \cite{Erasure}). Further, from Fig. \ref{osqc} we observe that in the range of frequencies in which $Q>0$ the curve with $\nu=10$ is larger than the one with $\nu=1000$. This suggests that, unlike the classical capacity, in order to maximize the quantum capacity with an impulsive accelerated mirror we are forced to take a final speed of the mirror comparable with the critical speed for high frequencies, i.e., $V\sim0.8$.

The minimum frequency required for a non-null quantum capacity depends upon $\nu$ through the relation
\begin{equation}
    \Omega^2=\frac{(1-\sqrt{\nu})^2}{2\nu}+\sqrt{\frac{(1-\sqrt{\nu})^4}{4\nu^2}+\frac{1}{\nu}}.
\end{equation}

For both $\nu\to1$ and $\nu\to\infty$ the frequency converges to $\Omega=1$. By construction, we thus expect a minimum in the range $1<\nu<\infty$.

\section{Final remarks}\label{Sec6}

In this work, we studied partially reflecting accelerating mirrors finding general expressions for the Bogoliubov coefficients. This work has been motivated by analog models, called accelerated boundary correspondences, that describe the correspondence between the particle production from a null-shell of a collapsing black hole and the particle production from a perfectly reflecting accelerating mirror.

The natural extension to semi-transparency indicated potential signatures of new effects. Along this line, we have studied the trajectories in which the mirror satisfies a few physical conditions: in the past, it lies at rest and in the future it shows a finite acceleration period, ending with a constant sub-light speed. We introduced the concept of impulsive accelerated mirrors and we computed the Bogoliubov coefficients by considering a very short acceleration period. Consequently, we evaluated particle production from the so-obtained Bogoliubov coefficients, providing explicit analytical expressions dependent on frequency and on the final speed of the mirror.

The particles considered are non-interactive scalar particles, hence with coupling constant $\lambda\rightarrow0$. It was proved in Refs.\cite{Interaction1,Interaction2} that the interaction (even if infinitesimal) gives a non-negligible contribute to the particle production at the time $t\sim\lambda^{-1}$. Here we considered an error on the frequency $\Delta(\omega)=\epsilon\pi$. As a consequence, the particles produced are the ones in the time interval $\left(-(\epsilon)^{-1},+(\epsilon)^{-1}\right)$. Since we have taken the continuous limit $\epsilon\to0$, in order to neglect the contribute of the interaction, it is sufficient to have $\lambda\rightarrow0$ faster than $\epsilon$. Moreover, considering mirrors with a finite acceleration period, we expect the particle production (and its consequent effects) to occur during the acceleration period. Hence, even if we consider $\lambda$ finite but very small, we do not care about what happens at times $t>\lambda^{-1}$, since the acceleration is over at such times.

Next, we have recognized the mirror as a Gaussian quantum channel acting between the spacetime regions of left-past and right-future. The evolution of an input signal crossing the mirror could then be studied using the previously obtained Bogoliubov coefficients.  For these quantum channels, we investigated the properties of transmission of an input signal, the noise created by the mirror over the channel and we finally evaluated both the classical and quantum capacities. Since we were searching for analytic solutions for the Bogoliubov coefficients, the continuous limit for the frequencies was considered. As a consequence, all the properties of the mirror as a Gaussian channel (i.e., $\tau$, $\overline{n}$ and the capacities) are an average in time from $-\infty$ to $+\infty$.

In addition, we speculated about the physical consequences of our framework. In particular, the simplicity and flexibility of the moving mirror model, coupled to its unique collection of radiative properties, demonstrate that with use of appropriate trajectories the moving mirror idealization of evaporating black hole radiation and information transfer are remarkably suitable. In harmony with black hole complementarity \cite{Susskind:1993if}, observers on both sides of the mirror cannot make simultaneous physical measurements, much the same way that one cannot both simultaneously measure, to within the uncertainty principle, the position and momentum of a particle in quantum mechanics.  With non-horizon perfect reflection, the information stays on one side of the mirror, carried by the radiation providing full knowledge of the initial quantum state; this necessarily requires that the radiation is never precisely thermal but quasi-thermal \cite{Good:2019tnf}.  That is, the particles are not distributed in an exact Planck distribution but carry small imprinting evidence of collapse. With non-horizon semi-transparency (e.g. Eq.~(\ref{pacc})), the right (left) observer collects the information from both the right (left)-movers which reflect (transmit) through the mirror, giving complete information about the initial state. For the sake of completeness, one can also consider the complementary communication scheme, namely the reflection case of an input signal incoming from right-past and outgoing to right-future. This prescription  is  likely less interesting than the transmission case. Indeed, the former works better in  modeling  black holes and information theory and holds a more appropriate physical meaning.
From the results of this work, we speculate that this picture in general, and further use of this class of trajectories in particular, will contribute towards the resolution of information transfer in the black hole evaporation process.

Concluding, to get relevant information about the time in which such properties occur, future works will generalize our treatment considering wave packets, whose wave packet width,  $\Delta\omega$, satisfies  $\Delta\omega\sim u_0^{-1}$ and so we will investigate physical properties during this small acceleration period, i.e., $u_0$.
The results lead smoothly to further investigations with respect to astrophysical applications to compact objects. For example, possible scenarios of high-energy astrophysical explosions could be object of future works modeled by means of our approach.

\acknowledgments
Funding from state-targeted program ``Center of Excellence for Fundamental and Applied Physics" (BR05236454) by the Ministry of Education and Science of the Republic of Kazakhstan is acknowledged, as well as the FY2021-SGP-1-STMM Faculty Development Competitive Research Grant No. 021220FD3951 at Nazarbayev University. OL is also thankful to the Ministry of Education and Science of the Republic of Kazakhstan, Grant: IRN AP08052311 for financial support.  SM is thankful to the funding from European Union's Horizon 2020 research and innovation program under grant agreement no. 862644 (FET-Open Project: QUARTET).

\end{document}